\documentclass[prl, longbibliography, aps, nofootinbib, twocolumn, amssymb, superscriptaddress, 10pt]{revtex4-2}
\usepackage{amsmath}
\usepackage{amssymb}
\usepackage{amsthm}
\usepackage{amsfonts}
\usepackage{enumerate}
\usepackage{latexsym}
\usepackage{color}
\usepackage{setspace} 
\usepackage{blindtext}
\usepackage{dsfont}
\usepackage{mathrsfs}
\usepackage[normalem]{ulem}
\usepackage[dvipsnames]{xcolor}

\usepackage{tabularx}
\newcolumntype{Y}{>{\centering\arraybackslash}X}

\usepackage{stackengine}

\usepackage{bm}
\usepackage{graphicx}

\usepackage{hyperref}
\hypersetup{
pdfnewwindow=true, colorlinks=true,
linkcolor=blue, anchorcolor=blue,
citecolor=blue, filecolor=blue,
menucolor=blue, urlcolor=blue} 

\usepackage{pifont}% http://ctan.org/pkg/pifont

\newcommand{\beginsupplement}{
        \setcounter{table}{0}
        \renewcommand{\thetable}{S\arabic{table}}
        \setcounter{figure}{0}
        \renewcommand{\thefigure}{S\arabic{figure}}
        \setcounter{equation}{0}
        \renewcommand{\theequation}{S\arabic{equation}}
        \setcounter{section}{0}
        \renewcommand{\thesection}{\Alph{section}}
        \setcounter{subsection}{0}
        \renewcommand{\thesubsection}{\arabic{subsection}}
}

\newcommand{\vk}{{\mathbf{k}}}
\newcommand{\vK}{{\mathbf{K}}}
\newcommand{\vp}{{\mathbf{p}}}
\newcommand{\vq}{{\mathbf{q}}}

\begin{document}

\title{Superconductivity in twisted \texorpdfstring{WSe$_2$}{WSe2} from topology-induced
quantum fluctuations}
\author{Fang Xie}
\thanks{\href{fx7@rice.edu}{fx7@rice.edu}}
\affiliation{Department of Physics \& Astronomy,  Extreme Quantum Materials Alliance, Smalley-Curl Institute, Rice University, Houston, Texas 77005, USA}
\affiliation{Rice Academy of Fellows, Rice University, Houston, Texas 77005, USA}
\author{Lei Chen}
\affiliation{Department of Physics \& Astronomy,  Extreme Quantum Materials Alliance, Smalley-Curl Institute,
Rice University, Houston, Texas 77005, USA}
\author{Shouvik Sur}
\affiliation{Department of Physics \& Astronomy,  Extreme Quantum Materials Alliance, Smalley-Curl Institute,
Rice University, Houston, Texas 77005, USA}
\author{Yuan Fang}
\affiliation{Department of Physics \& Astronomy,  Extreme Quantum Materials Alliance, Smalley-Curl Institute,
Rice University, Houston, Texas 77005, USA}
\author{Jennifer Cano}
\thanks{\href{jennifer.cano@stonybrook.edu}{jennifer.cano@stonybrook.edu}}
\affiliation{Department of Physics and Astronomy, Stony Brook University, Stony Brook, NY 11794, USA}
\affiliation{Center for Computational Quantum Physics, Flatiron Institute, New York, NY 10010, USA}
\author{Qimiao Si}
\thanks{\href{qmsi@rice.edu}{qmsi@rice.edu}}
\affiliation{Department of Physics \& Astronomy,  Extreme Quantum Materials Alliance, Smalley-Curl Institute, Rice University, Houston, Texas 77005, USA}

\date{\today}

\begin{abstract}
Recently, superconductivity has been observed in twisted WSe$_2$ moir\'{e} structures (\href{https://doi.org/10.1038/s41586-024-08116-2}{Xia et al., Nature 2024}; \href{https://doi.org/10.1038/s41586-024-08381-1}{Guo et al., Nature 2025}). Its transition temperature is high, reaching a few percent of the Fermi temperature scale. 
Here, we advance a mechanism for superconductivity based on the notion that electronic topology enables quantum fluctuations in a suitable regime of intermediate correlations. 
In this regime, the Coulomb interaction requires that an active topological flat band and nearby wider bands are considered together.
Compact molecular orbitals arise, which give rise to quantum fluctuations through topology-dictated hybridization with the other molecular orbitals. 
The hybridization competes with the active flat band's natural tendency towards static electronic ordering, thereby weakening the latter;
we link this effect with certain salient observations by experiments. 
Furthermore, the competition yields a quantum critical regime where quasiparticles are lost.
The corresponding quantum critical fluctuations drive superconductivity.
Broader implications and new connections among correlated materials platforms are discussed.
\end{abstract}

\maketitle

{\it
Introduction~~}
Moir\'{e} materials represent a rich platform to realize a plethora of strong correlation phenomena, spanning over Chern insulators, strange metals and superconductors \cite{cao2018correlated,cao2018unconventional,regan2020mott,tang2020simulation,wang2020correlated,xu2020correlated, li2021quantum,cai2023signatures,zeng2023thermodynamic,park2023observation, xu2023observation}. 
In these materials, exemplified by twisted bilayer graphene and transition metal dichalcogenides (TMDCs), both band topology and electronic correlations play key roles 
in organizing the phase diagram.
These features, in combination with the high tunability of the moir\'{e} potential and carrier concentration, have led to a practical platform for not only realizing  phenomena reminiscent of other classes of strongly correlated systems, such as heavy fermion compounds and cuprates~\cite{Keimer2017,Pas21.1}, but also accessing novel types of emergent behaviors~\cite{andrei2020graphene}.

Although superconductivity was identified early on in twisted bilayer graphene, only very recently has it been observed in any moir\'e TMDCs~\cite{Xia2024Unconventional, Guo2024Superconductivity}. 
Interestingly, the superconducting state in these twisted homo-bilayer WSe$_2$ (tWSe$_2$) systems is realized close to half-hole-filling $\nu \approx 1$.
It is characterized by a critical temperature $T_c$ of a few percent of $T_F$, the Fermi temperature, and a coherence length $\xi \sim 10  a_M$ with $a_M$ being the moir\'{e} lattice constant.
Both features are similar to that of typical superconducting states realized in correlated systems such as high-$T_c$ cuprates and heavy fermion compounds, and indicative of a strong pairing regime~\cite{Lee-RMP06,Si10.2,Si16.1}.
Moreover, depending on the twist angle, the superconducting state was observed to lie in the vicinity of a correlated phase~\cite{Xia2024Unconventional,Guo2024Superconductivity}.
For a flat band whose partial filling is commensurate, realizing a non-superconducting correlated state is not too surprising. What is striking is that tuning non-thermal parameters such as the displacement field can readily suppress this state in favor of the superconductor. 
This linkage also provides strong evidence for the unconventional nature of superconductivity.
In contrast, in moir\'e graphene systems, it is still a subject of debate as to whether the observed superconductivity is unconventional~\cite{wu2018theory, isobe2018unconventional, guo2018pairing, xu2018topological}.
The observed superconductivity in tWSe$_2$ is of considerable theoretical interest~\cite{Debanjan2024, Schrade2021Nematic,Chen2023Singlet,Zhou2023Chiral,Christos2024Approximate}.
However, several key issues have been opened up by the experiments, including not only the overall mechanism but also the question of why $T_c$ as measured by $T_F$ is as large as the experimental observation.

Here, we address these issues based on a combination of microscopic considerations and, in the process, advance a route to superconductivity that may be of general relevance to topological electronic systems.
Continuum models with parameters fitted to density functional theory calculations indicate 
that the bandwidth of the active bands in tWSe$_2$ is greatly suppressed 
by the moir\'{e} potential~\cite{Wu2019topological, Devakul2021Magic, Reddy2023Fractional}.
[See below, and the Supplemental Material (SM) Sec.~A \cite{supplemental_material}.]
With a partial filling of such a flat band, especially when it is commensurate, an electronic order is natural, and this feature is in line with the observation of the non-superconducting electron order in tWSe$_2$~\cite{Xia2024Unconventional,Guo2024Superconductivity}. To proceed, we are guided by two considerations. First, the flat band is expected to be topological. 
Second, the Coulomb interaction strength is expected to be larger than both the width of the flat band and its gap to wider bands (see below); accordingly, it requires that the flat and wider bands are examined together.
This is analogous to the case of kagome metals with active flat and wide bands, where the flat bands are represented by compact molecular orbitals \cite{hu2023coupled,chen2024emergent, chen2023metallic}, leading to a description via topological Kondo lattice models~\cite{lai2018weyl,dzsaber2021giant}.
Importantly, as dictated by band topology, the compact molecular orbitals {\it must hybridize} with other orbitals.
The hybridization amounts to quantum fluctuations: the competition between this hybridization and the tendency of the active flat bands to form static electronic order leads to a quantum critical regime;
the qualitative physics, by analogy with that of the heavy fermion metals~\cite{Hu-Natphys2024} and supported by our calculations reported here, is captured by a schematic phase diagram [see below, Fig.~\ref{fig:phdg}(a)].
We study the model in terms of the extended-dynamical mean field theory (EDMFT) method because it is essential to treat the aforementioned competition in a dynamical way (instead of by the Hartree-Fock method)~\cite{Hu-Natphys2024,Hu2022extended}, and we use its cluster form so that unconventional pairing can be captured~\cite{Hu2021-sc}.

\begin{figure}[t]
    \centering
    \includegraphics[width=\linewidth]{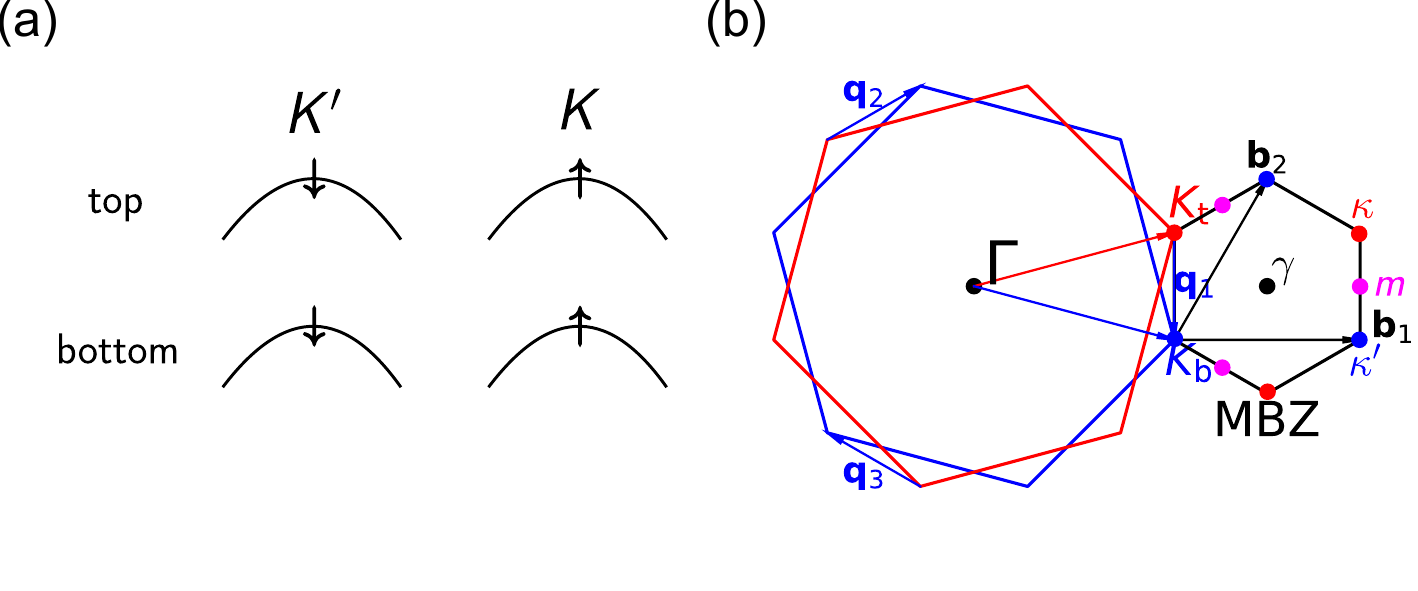}
    \caption{(a) The spin-valley locking mechanism of transition metal dichalcogenides in the $2H$ structural phase.
    (b) The moir\'e Brillouin zone (black hexagon) is spanned by the difference between the $K$ (or $K'$) points in the top and bottom layers.
    }
    \label{fig:models}
\end{figure}

{\it Model and method: Band structure and Wannier functions~~}
We start from the low-energy effective model of twisted bilayer WSe$_2$ in the $2H$ structural phase. The single-layer band structure around $K$ and $K'$ points in these TMDC materials can be approximately described by a hole pocket with quadratic dispersion.
The two hole pockets around the two valleys have opposite spin orientation as shown in Fig.~\ref{fig:models}(a), which is referred to as spin-valley locking~\cite{Liu2013Three,Kormanyos2015kp}.
When the two layers of $\rm WSe_2$ are stacked together with a small twisting angle $\theta$, the single layer Brillouin zones will have a mismatch, leading to a moir\'e superlattice. 
The moir\'e Brillouin zone is spanned by the difference between the $K$ (or $K'$) points in the top and bottom layers, as shown in Fig.~\ref{fig:models}(b), in which the reciprocal basis vectors of the moir\'e Brillouin zone is labeled by $\mathbf{b}_1$ and $\mathbf{b}_2$.
The Bravais lattice basis vectors $\mathbf{a}_{1,2}$ is determined accordingly.

\begin{figure}[t]
    \centering
    \includegraphics[width=\linewidth]{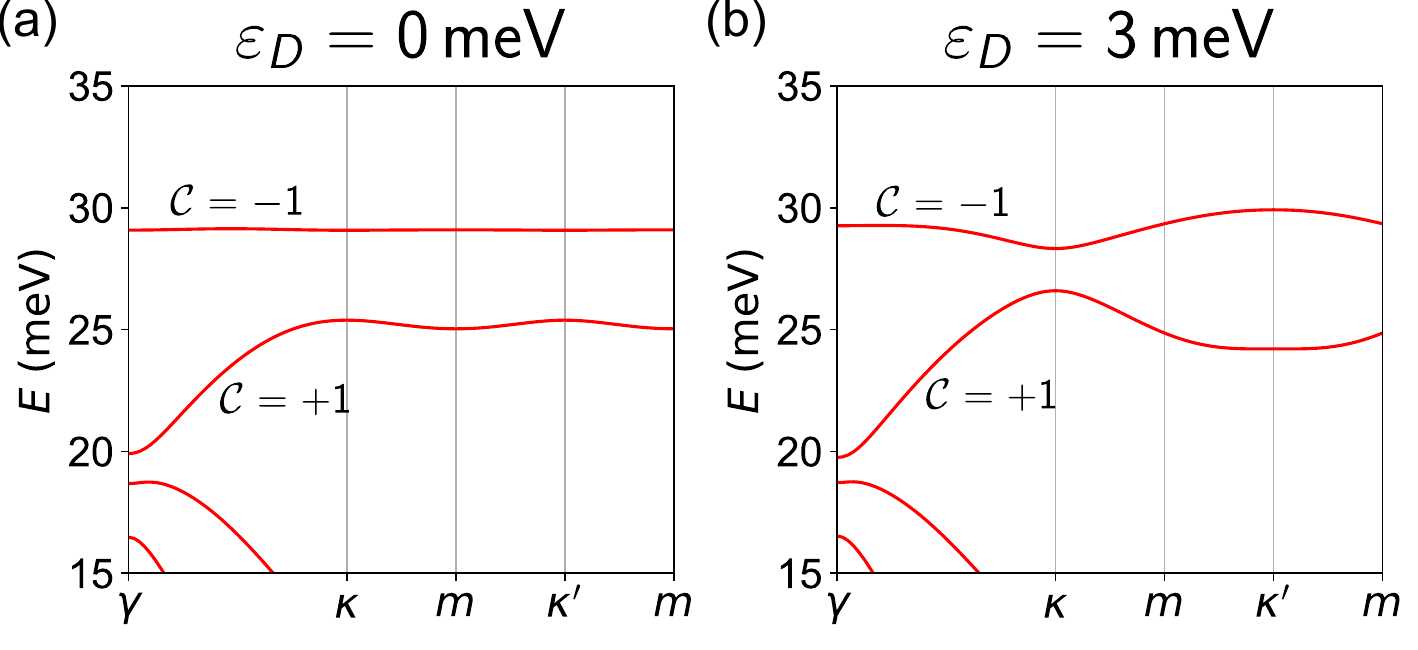}
    \caption{Single valley band structure of the continuum model with two different displacement field strengths $\varepsilon_D$. In both cases, the top two moir\'e bands carry opposite valley Chern numbers. The twisting angle is chosen to be $\theta = 1.43^\circ$ for illustration purpose; 
    similar considerations apply to larger twist angles (See Sec.~B of the SM \cite{supplemental_material}).
    }
    \label{fig:bm-bands}
\end{figure}

\begin{figure}[t]
    \centering
    \includegraphics[width=\linewidth]{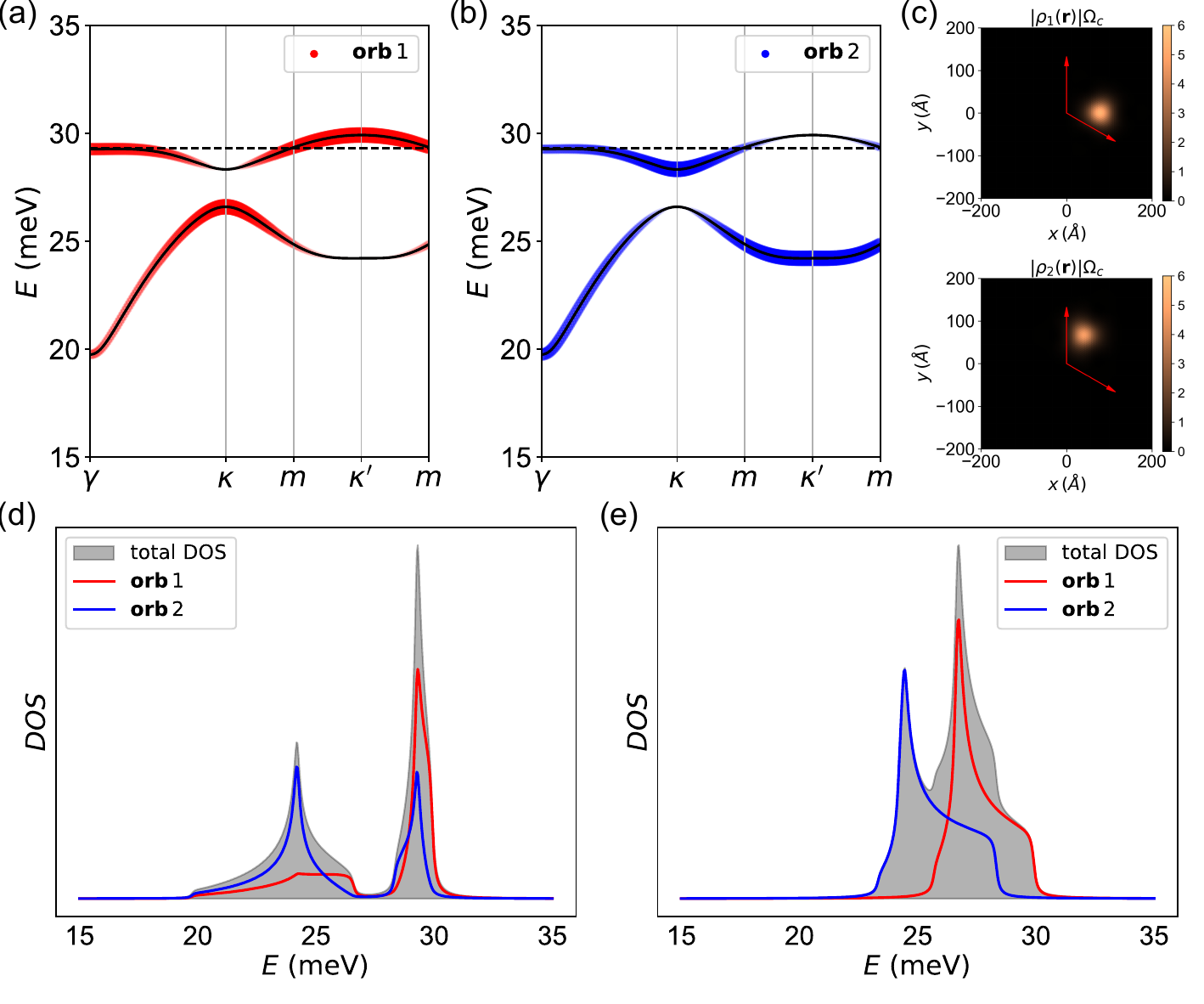}
    \caption{
        (a-b) The fat band projection of the two moir\'e orbitals with displacement field potential strength $\varepsilon_D = 3\,\rm meV$.
        The dashed lines stand for the Fermi level in the non-interacting band structure at hole filling factor $\nu = 1$.
        (c) The density distribution of the Wannier functions of the two moir\'e orbitals. The two red arrows stand for the Bravais lattice basis vectors.
        (d) The density of states of the two-orbital tight-binding model.
        (e) The density of states with the hybridization between the two orbitals removed.
    }
    \label{fig:orbitals}
\end{figure}

The continuum Hamiltonian, which is discussed in detail in Sec.~A of the SM \cite{supplemental_material}, can be constructed to compute the band structure.
In Fig.~\ref{fig:bm-bands}, we show the band structure of the continuum model with two different displacement field strengths $\varepsilon_D$, with the twisting angle $\theta = 1.43^\circ$;
a similar regime of band topology can also arise for larger twisting angles as illustrated in Sec.~B of the SM \cite{supplemental_material}.
Symmetry analysis of the moir\'e bands leads to a two-orbital tight-binding model, which faithfully describe the two active bands of the continuum model in a single valley, and the detail of the tight-binding parameters are provided in Sec.~C of the SM \cite{supplemental_material}.
Each of the two oritals forms a set of triangular lattice, and they together form a hexagonal lattice.
The projection of the two moir\'e orbitals onto the top bands, and the charge density distributions of them are shown in Figs.~\ref{fig:orbitals}(a-c).
In Fig.~\ref{fig:orbitals}(d), we also provide the density of states, in which the first orbital has a larger weight in the top band, while the second orbital has a larger weight in the second band.
As already emphasized, the topological nature of the bands ensure that the two orbitals hybridize with each other.
In fact, the nearest-neighbor hybridization between the two orbitals is noticeably larger than the intra-orbital hopping terms.
In addition, the on-site energies of these two orbitals are also different in the presence of a displacement field, which could give rise to very different relative fillings of these two orbitals at total hole filling $\nu = 1$, and lead to orbital-selective correlations, with those closer to half-filling experiencing stronger correlation effects \cite{Si16.1}.
This on-site energy difference is clearly evident in the DOS plot shown in Fig.~\ref{fig:orbitals}(e), where the hybridization terms are turned off, and the density of states distributions of the two orbitals are noticeably separated.

{\it Model and method: Interaction Hamiltonian and extended-dynamical mean field theory}~~
We implicitly assume that the Coulomb interaction $U$ is larger than the width of the top hole band but does not exceed the overall band widths of the two top bands. What happens when $U$ is further reduced or increased is described in some detail in the SM (Sec.~F), where the regime of our interest is marked as regime iii in Fig.~S2. 
In this regime, we can utilize the two-orbital tight-binding model to construct the interaction Hamiltonian.
Numerically performing the direct channel overlap integral of the double-gate screened Coulomb interaction, we find that the on-site interaction $U$ is on the order of magnitude of $10\sim30\,\rm meV$.
The estimated $U$ depends on the gating setup and dielectric constants of the substrate and the sample; a detailed discussion is provided in SM Sec.~D \cite{supplemental_material}.
In addition, due to its slightly smaller orbital size, the on-site interaction of the orbital 1 is slightly larger than that of the orbital 2.
Furthermore, we find that the nearest-neighbor interaction is less than $20\%$ of the on-site interaction, and all the exchange interactions are less than $5\%$ of $U$, so the Hubbard terms are dominant \cite{Wu2018Hubbard}.
Thus, the effective interaction Hamiltonian can be written as:
\begin{equation}
    H_I = U\sum_{\mathbf{R}_0\alpha} n_{\mathbf{R}_0,\alpha,\uparrow}n_{\mathbf{R}_0,\alpha,\downarrow}\,,
\end{equation}
in which $\alpha,\beta \in \{1,2\}$ stand for the two orbitals.

In practice, orbital 1 has a hole occupancy ($\approx 0.8$) that is considerably closer to half-filling than orbital 2 (which has a hole occupancy $\approx 0.2$) in the non-interacting band structure. 
Given that its filling factor is much closer to half-filling, the interaction effects in orbital 1 are expected to be significantly stronger than those in orbital 2 \cite{fazekas1999lecture, Kotliar1986, Hassan2010slave}, and therefore favor being more localized. A simple explanation of this behavior, based on a single-band Hubbard model, is provided in Sec.~F of the SM \cite{supplemental_material}.
We will denote orbitals $1$ and $2$ as 
the more localized $f$ and conduction electron $c$ orbitals, respectively, and focus our treatment on the correlation effects of the $f$ orbital. This leads to an effective periodic Anderson Hamiltonian, which takes the form:
\begin{align}
    H_{\mathrm{AL}} =& \sum_{\vp, \sigma} (\epsilon_{\vp} -\mu) c_{\vp\sigma}^{\dag}c_{\vp\sigma} + \sum_{i} \left[(\epsilon_f-\mu) n_{fi} +U n_{fi\uparrow}n_{fi\downarrow}\right]\nonumber\\ 
    &+\sum_{\vp, i\sigma}\left(V_{\vp i} c_{\vp\sigma}^{\dag}f_{i\sigma} + \mathrm{h.c.} \right) + \sum_{i,j,m} I_{ij}S^{m}_{fi} S^{m}_{fj} \,,
\label{eqn:ham}
\end{align}
where $c_{\vp\sigma}$ ($f_{i \sigma}$) annihilates a $c$ ($f$) electron at momentum $\vp$ (lattice site $i$) with spin $\sigma$, $\epsilon_{\vp}$ is the dispersion of the conduction electrons while $\epsilon_f$ and $U$ denote the $f$-level energy and on-site Coulomb repulsion, respectively. 
In addition, $\mu$ is the chemical potential.
To incorporate its quantum fluctuations and facilitate the many-body calculations described below, it is adequate to treat the hybridization between the $f$ and $c$ electrons at an average level, setting $V_{\vp i} = V_{\rm hyb}$. 
We use the following estimated parameters: $\epsilon_f = 27.60\, \rm meV$, $V_{\rm hyb} = 2.5\, \rm meV$, $\mu=27.67\,\rm meV$ and $U= 10\, \rm meV$. 
Here, the value of on-site energy $\epsilon_f$ is directly obtained from the effective tight-binding model, the chemical potential $\mu$ is determined by the hole filling factor $\nu = 1$, and the hybridization is estimated by the intra-orbital hopping strength, which is discussed in detail in SM Sec.~C \cite{supplemental_material}.
The bare conduction electron spectrum $\epsilon_{\vp}$ is derived from the tight-binding parameters, with its corresponding DOS shown in Fig.~\ref{fig:orbitals}(e).
$I_{ij}$ represents the emergent Ruderman–Kittel–Kasuya–Yosida (RKKY) exchange interactions, mediated by the conduction $c$ electrons, for the Cartesian component $m \in \{x, y, z\}$ of the $f$ electrons.

We study the effective model using the EDMFT method~\cite{Hu-Natphys2024,Hu2022extended}.
To facilitate the investigation of unconventional superconducting pairing, we utilize a cluster version of the EDMFT (C-EDMFT) approach \cite{Hu2021-sc,Pixley-c-edmft2015}, which determines correlation functions of $H_{\rm AL}$ in terms of a self-consistently determined two-site quantum cluster model that involves both fermionic and bosonic baths. 
The numerically exact continuous-time quantum Monte Carlo (CTQMC) method is used to solve the effective cluster model at nonzero temperatures. 
Multiple physical quantities, including the expectation value of the spin operator, and the pairing susceptibility can be evaluated.
The technical detail of C-EDMFT is also discussed in the SM (Sec.~E) \cite{supplemental_material}.
The transition temperature for the ordering state and the superconducting states can therefore be determined by the non-zero spin expectation and divergence of pairing susceptibility at different temperatures.
In the simulation, we also treat the Kondo energy scale as a fixed value and directly tune the value of $I$ as an independent variable, which is more convenient in the EDMFT simulation. 
This approach is justified by the expectation that the phase diagram with quantum critical point induced by the competition between Kondo screening and RKKY exchange is universal.

{\it Quantum fluctuations and superconductivity~~}
Our mapping establishes the connection between tWSe$_2$ and canonical heavy fermion systems, where quantum fluctuations have been proven to play a crucial role, giving rise to 
such phenomena as metallic quantum criticality and unconventional superconductivity~\cite{Hu2021-sc}. 
The new aspect of the present system is that the quantum fluctuations are driven by the topologically-required hybridization between the localized and extended orbitals.

\begin{figure}[t!]
    \centering
    \includegraphics[width=\linewidth]{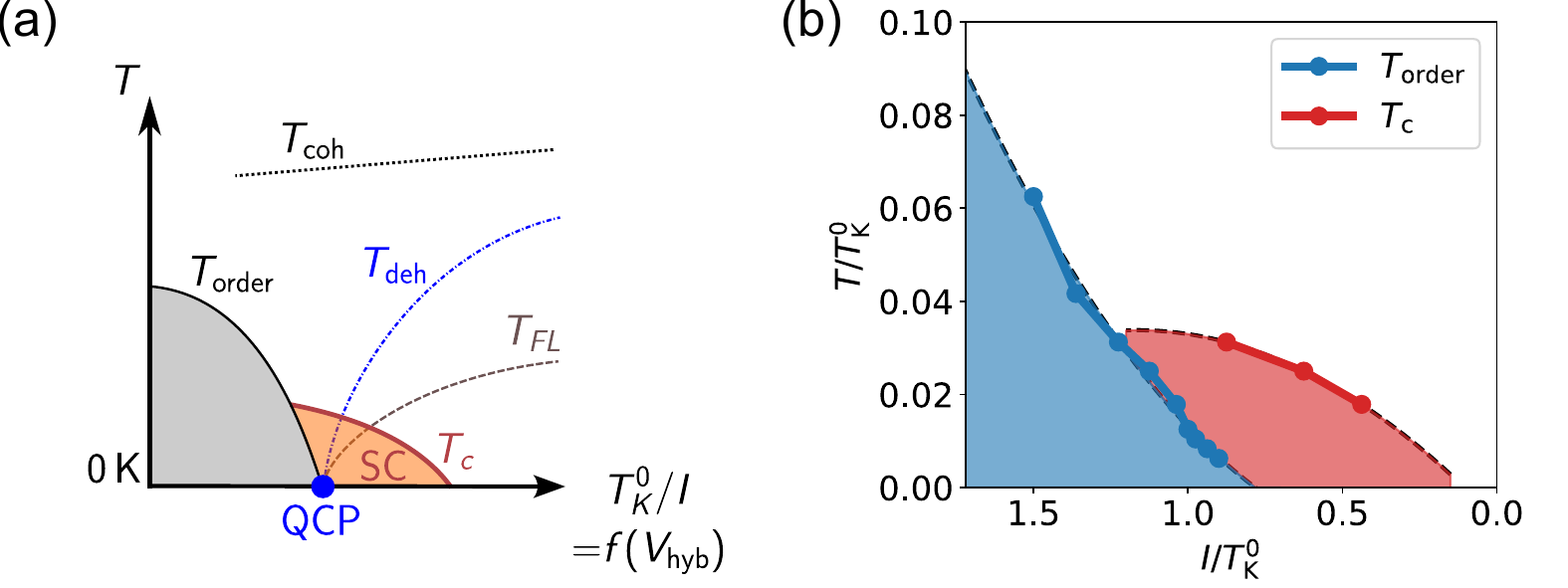}
    \caption{(a) Schematic phase diagram in temperature and as a function of the hybridization strength, $V_{\rm hyb}$, for a fixed set of interaction and other coupling parameters. 
    Tuning $V_{\rm hyb}$ can effectively control the ratio between Kondo screening and the RKKY interaction, by which a quantum critical point can be achieved.
    It shows the vanishing of the dehybridization energy scale ($T_{\rm deh}$)~\cite{Hu-Natphys2024,Si01.1} at the QCP.
    Here $T_{\rm order}$ and $T_{\rm c}$ are the transition temperatures to an electronic order and to superconducting state (SC), respectively. 
    $T_{\rm FL}$ is a low-temperature crossover scale into a Fermi liquid, and $T_{\rm coh}$ a high-temperature crossover scale for the initial onset of Kondo coherence.
    (b) Finite-temperature phase diagram as a function of the control parameter $I/T_{K}^{0}$ (which is tuned by the hybridization) obtained from C-EDMFT calculations, with $T_{\rm c}$ and $T_{\rm order}$ marked by red and blue circles, respectively. 
    }
    \label{fig:phdg}
\end{figure}

Turning to our results, we start by showing in Fig.~\ref{fig:phdg}(a) a schematic phase diagram when the hybridization strength $V_{\rm hyb}$ is tuned.
This parameter implicitly controls the RKKY interaction strength $I \sim \rho_0J_K^2$ via the effective Kondo coupling $J_K \sim V_{\rm hyb}^2\left(\frac{1}{\epsilon_f}+\frac{1}{-\epsilon_f + U}\right)$, where $\rho_0$ is the conduction band density of states.
The relative strengths of the Kondo temperature scale, $T^0_K \sim \rho^{-1}_0 e^{-1/(J_K \rho_0)}$ , which measures the degree of screening of local spin moments, and the RKKY interaction $I$, which drives the formation of symmetry-breaking ordered states, are governed by the hybridization strength. This competition ultimately tunes the system across metallic quantum criticality.
In the SM (Sec.~F) \cite{supplemental_material}, we discuss the correlated phases in further detail.

These expectations are supported by our numerical results, which are shown in Fig.~\ref{fig:phdg}(b). 
When the RKKY interaction dominates, the system develops an antiferromagnetic order of the locked spin-valley degree of freedom. As already emphasized, because the chemical potential passes through the flat band [{\it c.f.} Fig.~\ref{fig:orbitals}(a)], this electronic ordering is natural \cite{Mielke_1992, Lin2023Complex}.
The hybridization promotes the Kondo effect; increasing it amounts to decreasing the ratio $I/T_{\rm K}^0$ [going rightward in both Figs.~\ref{fig:phdg}(a-b)].
It weakens the order, eventually suppressing it and reaching a QCP, which in turn drives the superconductivity presented in Fig.~\ref{fig:phdg}(b).

We calculate the static lattice pairing susceptibility~\cite{Hu2021-sc}, whose divergence identifies the superconducting transition temperature $T_c$. We perform these calculations at various temperatures to determine the finite-temperature phase boundary for the superconducting phase.
The resulting phase diagram is displayed in Fig.~\ref{fig:phdg}(b), with the temperatures measured by the bare Kondo temperature ($T_{\rm K}^0$), which serves as the effective Fermi temperature. 
The region of superconducting order is anchored by the underlying QCP.
The superconducting transition temperature $T_c$ near the QCP reaches approximately 3$\%$ of the Fermi temperature.

To further connect our numerical results with the schematic phase diagram Fig.~\ref{fig:phdg}(a), we note that, at the QCP, not only does the transition from order to disorder take place, but a new energy scale, $T_{\rm deh}$ --- which characterizes the hybridization-dehybridization between the $f$ and $c$ electrons --- also vanishes~\cite{Hu-Natphys2024,Si01.1,Pas21.1,Kirchner_RMP,Pas04.1}, indicating a critical destruction of Landau quasiparticles. 
The superconducting phase emerges near the QCP, where the quantum fluctuations are the most pronounced.

Finally, to address the robustness of this superconducting mechanism against parameter changes, we also performed the C-EDMFT calculation for another set of parameters with different interaction strengths. The results can be found in Sec.~G of the SM \cite{supplemental_material}. The combination of Figs.~S5(a-b) and Fig.~\ref{fig:phdg}(b) allows us to conclude that the proposed mechanism, viz. superconductivity driven by the topology-induced quantum fluctuations near the QCP, can emerge over an extended regime of model parameters. In other words, this mechanism is robust.

{\it Discussion and summary~~}
Several remarks are in order.
Firstly, the conceptual novelty of our work is to emphasize that band topology amounts to quantum fluctuations in the suitable regime of intermediate correlations (see the SM: Fig.~S2, regime iii).
The quantum fluctuations appear through hybridization. 
tWSe$_2$ provides an ideal setting to realize this effect. In the underlying bands illustrated by Fig.~\ref{fig:orbitals}(a), in the absence of band topology, the real space orbitals would have been individually representing the two bands and would be kinetically decoupled from each other. Topology dictates that these real space orbitals are hybridized with each other, which competes with the flat-band-system's natural tendency towards non-superconducting electronic order.

Secondly, for the correlated phase we have emphasized the electronic order on a triangular lattice.
At commensurate filling, this phase is naturally insulating, as is observed experimentally~\cite{Xia2024Unconventional}. 
In contrast, away from commensurate filling, the system would stay metallic, which also is seen experimentally~\cite{Guo2024Superconductivity}. 

Thirdly, our work highlights the role of topology-enabled quantum fluctuations 
and how the resulting quantum criticality drives the superconductivity [cf. Fig.~\ref{fig:phdg}(a)].
While the experimental results in tWSe$_2$ are still limited, evidence for the influence of quantum criticality has already emerged. For example, a canonical means to empirically assess the proximity to quantum criticality is to see how strongly the resistivity $A$-coefficient varies with a control parameter, as demonstrated in quantum critical heavy fermion metals~\cite{Pas21.1}.
Here, $A$ is the coefficient of the $T^2$ dependence in the electrical resistivity. In tWSe$_2$, $A^{1/2}$ is found to increase by more than a factor of $10$ as the filling moves towards the normal state of the superconducting phase~\cite{Xia2024Unconventional},
providing the evidence for the quantum critical mechanism for superconductivity advanced in this work.
We elaborate our discussion regarding the experimental evidence for quantum criticality in the Sec.~H of the SM \cite{supplemental_material}. 
We note in passing that related Kondo lattice behavior in hetero-layers of TMDCs has been observed experimentally \cite{zhao2022gate, zhao2023emergence} and discussed theoretically \cite{Guerci2023Chiral, Xie2024Kondo}. 

Fourthly, our work reveals a description of the tWSe$_2$ system in terms of a topological Kondo lattice. 
This is expected to give rise to Kondo like behavior in the temperature dependence of the electrical resistivity, which can be most clearly 
captured in the regime near the coherence temperature. 
Such behavior is indeed seen experimentally~\cite{Xia2024Unconventional}. 
At the same time, in the absence of the displacement field, the $ C_{2y}T$ symmetry would render the Kondo lattice to a two-orbital extended Hubbard model with inter-orbital hybridization. 
We expect a smooth crossover of the physics between the two regimes.

Finally, our results capture the high superconducting transition temperature 
in the sense that $T_c$ is about a few percent of the Fermi temperature. 
This feature is shared by such bulk correlated materials as the heavy fermion metals and cuprates~\cite{Hu2021-sc,Hu-Natphys2024}. 
As such, our work reveals new connections between the unconventional superconductivity across the platforms for strongly correlated systems.
This interconnection suggests that the mechanism we have advanced may operate in other correlated systems, including other moir\'e structures~\cite{Ram2021,Song2022,Kumar2022, Datta2023Heavy} and kagome and pyrochlore systems~\cite{hu2023coupled,chen2024emergent,chen2023metallic,Huang2023np} that connect with heavy fermion systems.

To summarize, we advance a mechanism for superconductivity recently discovered in the twisted WSe$_2$ moir\'e structures and provide the understanding for the observed superconducting transition temperature that is as high as a few percent of the Fermi temperature.
We show how the electronic topology in this system enables quantum fluctuations, which gives rise to quantum criticality and the concomitant superconductivity. 
More generally, our results suggest that, in suitable regimes of correlations, band topology amounts to quantum fluctuations. 
Furthermore, our work reveals new connections among a variety of strongly correlated systems.

\begin{acknowledgments}
We thank Valentin Cr\'epel, Zhongdong Han, Andrew Millis, Abhay Pasupathy, Silke Paschen, Yiyu Xia and Yonglong Xie and, 
particularly Kin Fai Mak and Jie Shan for useful discussions.
This work has been supported in part by the NSF Grant No.\ DMR-2220603 (F.X. and L.C.), the AFOSR under 
Grant No.\ FA9550-21-1-0356 (S.S.,Y.F.), the Robert A. Welch Foundation Grant No.\ C-1411 (Q.S.) and 
the Vannevar Bush Faculty Fellowship ONR-VB N00014-23-1-2870 (Q.S.).
J.C. acknowledges the support of the National Science Foundation under Grant No. DMR-1942447, support from the Alfred P. Sloan Foundation through a Sloan Research Fellowship and the support of the Flatiron Institute, a division of the Simons Foundation.
The majority of the computational calculations have been performed on the Shared University Grid at Rice funded by NSF under Grant No.~EIA-0216467, a partnership between Rice University, Sun Microsystems, and Sigma Solutions, Inc., the Big-Data Private-Cloud Research Cyberinfrastructure MRI-award funded by NSF under Grant No. CNS-1338099, and the Extreme Science and Engineering Discovery Environment (XSEDE) by NSF under Grant No. DMR170109. 
Q.S. acknowledges the hospitality of the Aspen Center for Physics, which is supported by NSF grant No. PHY-2210452.

\end{acknowledgments}

\bibliography{reference.bib}

\begin{thebibliography}{66}%
\makeatletter
\providecommand \@ifxundefined [1]{%
 \@ifx{#1\undefined}
}%
\providecommand \@ifnum [1]{%
 \ifnum #1\expandafter \@firstoftwo
 \else \expandafter \@secondoftwo
 \fi
}%
\providecommand \@ifx [1]{%
 \ifx #1\expandafter \@firstoftwo
 \else \expandafter \@secondoftwo
 \fi
}%
\providecommand \natexlab [1]{#1}%
\providecommand \enquote  [1]{``#1''}%
\providecommand \bibnamefont  [1]{#1}%
\providecommand \bibfnamefont [1]{#1}%
\providecommand \citenamefont [1]{#1}%
\providecommand \href@noop [0]{\@secondoftwo}%
\providecommand \href [0]{\begingroup \@sanitize@url \@href}%
\providecommand \@href[1]{\@@startlink{#1}\@@href}%
\providecommand \@@href[1]{\endgroup#1\@@endlink}%
\providecommand \@sanitize@url [0]{\catcode `\\12\catcode `\$12\catcode
  `\&12\catcode `\#12\catcode `\^12\catcode `\_12\catcode `\%12\relax}%
\providecommand \@@startlink[1]{}%
\providecommand \@@endlink[0]{}%
\providecommand \url  [0]{\begingroup\@sanitize@url \@url }%
\providecommand \@url [1]{\endgroup\@href {#1}{\urlprefix }}%
\providecommand \urlprefix  [0]{URL }%
\providecommand \Eprint [0]{\href }%
\providecommand \doibase [0]{https://doi.org/}%
\providecommand \selectlanguage [0]{\@gobble}%
\providecommand \bibinfo  [0]{\@secondoftwo}%
\providecommand \bibfield  [0]{\@secondoftwo}%
\providecommand \translation [1]{[#1]}%
\providecommand \BibitemOpen [0]{}%
\providecommand \bibitemStop [0]{}%
\providecommand \bibitemNoStop [0]{.\EOS\space}%
\providecommand \EOS [0]{\spacefactor3000\relax}%
\providecommand \BibitemShut  [1]{\csname bibitem#1\endcsname}%
\let\auto@bib@innerbib\@empty
%</preamble>
\bibitem [{\citenamefont {Cao}\ \emph {et~al.}(2018{\natexlab{a}})\citenamefont
  {Cao}, \citenamefont {Fatemi}, \citenamefont {Demir}, \citenamefont {Fang},
  \citenamefont {Tomarken}, \citenamefont {Luo}, \citenamefont
  {Sanchez-Yamagishi}, \citenamefont {Watanabe}, \citenamefont {Taniguchi},
  \citenamefont {Kaxiras} \emph {et~al.}}]{cao2018correlated}%
  \BibitemOpen
  \bibfield  {author} {\bibinfo {author} {\bibfnamefont {Y.}~\bibnamefont
  {Cao}}, \bibinfo {author} {\bibfnamefont {V.}~\bibnamefont {Fatemi}},
  \bibinfo {author} {\bibfnamefont {A.}~\bibnamefont {Demir}}, \bibinfo
  {author} {\bibfnamefont {S.}~\bibnamefont {Fang}}, \bibinfo {author}
  {\bibfnamefont {S.~L.}\ \bibnamefont {Tomarken}}, \bibinfo {author}
  {\bibfnamefont {J.~Y.}\ \bibnamefont {Luo}}, \bibinfo {author} {\bibfnamefont
  {J.~D.}\ \bibnamefont {Sanchez-Yamagishi}}, \bibinfo {author} {\bibfnamefont
  {K.}~\bibnamefont {Watanabe}}, \bibinfo {author} {\bibfnamefont
  {T.}~\bibnamefont {Taniguchi}}, \bibinfo {author} {\bibfnamefont
  {E.}~\bibnamefont {Kaxiras}}, \emph {et~al.},\ }\href
  {https://doi.org/10.1038/nature26154} {\bibfield  {journal} {\bibinfo
  {journal} {Nature}\ }\textbf {\bibinfo {volume} {556}},\ \bibinfo {pages}
  {80} (\bibinfo {year} {2018}{\natexlab{a}})}\BibitemShut {NoStop}%
\bibitem [{\citenamefont {Cao}\ \emph {et~al.}(2018{\natexlab{b}})\citenamefont
  {Cao}, \citenamefont {Fatemi}, \citenamefont {Fang}, \citenamefont
  {Watanabe}, \citenamefont {Taniguchi}, \citenamefont {Kaxiras},\ and\
  \citenamefont {Jarillo-Herrero}}]{cao2018unconventional}%
  \BibitemOpen
  \bibfield  {author} {\bibinfo {author} {\bibfnamefont {Y.}~\bibnamefont
  {Cao}}, \bibinfo {author} {\bibfnamefont {V.}~\bibnamefont {Fatemi}},
  \bibinfo {author} {\bibfnamefont {S.}~\bibnamefont {Fang}}, \bibinfo {author}
  {\bibfnamefont {K.}~\bibnamefont {Watanabe}}, \bibinfo {author}
  {\bibfnamefont {T.}~\bibnamefont {Taniguchi}}, \bibinfo {author}
  {\bibfnamefont {E.}~\bibnamefont {Kaxiras}},\ and\ \bibinfo {author}
  {\bibfnamefont {P.}~\bibnamefont {Jarillo-Herrero}},\ }\href
  {https://doi.org/10.1038/nature26160} {\bibfield  {journal} {\bibinfo
  {journal} {Nature}\ }\textbf {\bibinfo {volume} {556}},\ \bibinfo {pages}
  {43} (\bibinfo {year} {2018}{\natexlab{b}})}\BibitemShut {NoStop}%
\bibitem [{\citenamefont {Regan}\ \emph {et~al.}(2020)\citenamefont {Regan},
  \citenamefont {Wang}, \citenamefont {Jin}, \citenamefont {Bakti~Utama},
  \citenamefont {Gao}, \citenamefont {Wei}, \citenamefont {Zhao}, \citenamefont
  {Zhao}, \citenamefont {Zhang}, \citenamefont {Yumigeta} \emph
  {et~al.}}]{regan2020mott}%
  \BibitemOpen
  \bibfield  {author} {\bibinfo {author} {\bibfnamefont {E.~C.}\ \bibnamefont
  {Regan}}, \bibinfo {author} {\bibfnamefont {D.}~\bibnamefont {Wang}},
  \bibinfo {author} {\bibfnamefont {C.}~\bibnamefont {Jin}}, \bibinfo {author}
  {\bibfnamefont {M.~I.}\ \bibnamefont {Bakti~Utama}}, \bibinfo {author}
  {\bibfnamefont {B.}~\bibnamefont {Gao}}, \bibinfo {author} {\bibfnamefont
  {X.}~\bibnamefont {Wei}}, \bibinfo {author} {\bibfnamefont {S.}~\bibnamefont
  {Zhao}}, \bibinfo {author} {\bibfnamefont {W.}~\bibnamefont {Zhao}}, \bibinfo
  {author} {\bibfnamefont {Z.}~\bibnamefont {Zhang}}, \bibinfo {author}
  {\bibfnamefont {K.}~\bibnamefont {Yumigeta}}, \emph {et~al.},\ }\href
  {https://doi.org/10.1038/s41586-020-2092-4} {\bibfield  {journal} {\bibinfo
  {journal} {Nature}\ }\textbf {\bibinfo {volume} {579}},\ \bibinfo {pages}
  {359} (\bibinfo {year} {2020})}\BibitemShut {NoStop}%
\bibitem [{\citenamefont {Tang}\ \emph {et~al.}(2020)\citenamefont {Tang},
  \citenamefont {Li}, \citenamefont {Li}, \citenamefont {Xu}, \citenamefont
  {Liu}, \citenamefont {Barmak}, \citenamefont {Watanabe}, \citenamefont
  {Taniguchi}, \citenamefont {MacDonald}, \citenamefont {Shan} \emph
  {et~al.}}]{tang2020simulation}%
  \BibitemOpen
  \bibfield  {author} {\bibinfo {author} {\bibfnamefont {Y.}~\bibnamefont
  {Tang}}, \bibinfo {author} {\bibfnamefont {L.}~\bibnamefont {Li}}, \bibinfo
  {author} {\bibfnamefont {T.}~\bibnamefont {Li}}, \bibinfo {author}
  {\bibfnamefont {Y.}~\bibnamefont {Xu}}, \bibinfo {author} {\bibfnamefont
  {S.}~\bibnamefont {Liu}}, \bibinfo {author} {\bibfnamefont {K.}~\bibnamefont
  {Barmak}}, \bibinfo {author} {\bibfnamefont {K.}~\bibnamefont {Watanabe}},
  \bibinfo {author} {\bibfnamefont {T.}~\bibnamefont {Taniguchi}}, \bibinfo
  {author} {\bibfnamefont {A.~H.}\ \bibnamefont {MacDonald}}, \bibinfo {author}
  {\bibfnamefont {J.}~\bibnamefont {Shan}}, \emph {et~al.},\ }\href
  {https://doi.org/10.1038/s41586-020-2085-3} {\bibfield  {journal} {\bibinfo
  {journal} {Nature}\ }\textbf {\bibinfo {volume} {579}},\ \bibinfo {pages}
  {353} (\bibinfo {year} {2020})}\BibitemShut {NoStop}%
\bibitem [{\citenamefont {Wang}\ \emph {et~al.}(2020)\citenamefont {Wang},
  \citenamefont {Shih}, \citenamefont {Ghiotto}, \citenamefont {Xian},
  \citenamefont {Rhodes}, \citenamefont {Tan}, \citenamefont {Claassen},
  \citenamefont {Kennes}, \citenamefont {Bai}, \citenamefont {Kim} \emph
  {et~al.}}]{wang2020correlated}%
  \BibitemOpen
  \bibfield  {author} {\bibinfo {author} {\bibfnamefont {L.}~\bibnamefont
  {Wang}}, \bibinfo {author} {\bibfnamefont {E.-M.}\ \bibnamefont {Shih}},
  \bibinfo {author} {\bibfnamefont {A.}~\bibnamefont {Ghiotto}}, \bibinfo
  {author} {\bibfnamefont {L.}~\bibnamefont {Xian}}, \bibinfo {author}
  {\bibfnamefont {D.~A.}\ \bibnamefont {Rhodes}}, \bibinfo {author}
  {\bibfnamefont {C.}~\bibnamefont {Tan}}, \bibinfo {author} {\bibfnamefont
  {M.}~\bibnamefont {Claassen}}, \bibinfo {author} {\bibfnamefont {D.~M.}\
  \bibnamefont {Kennes}}, \bibinfo {author} {\bibfnamefont {Y.}~\bibnamefont
  {Bai}}, \bibinfo {author} {\bibfnamefont {B.}~\bibnamefont {Kim}}, \emph
  {et~al.},\ }\href {https://doi.org/10.1038/s41563-020-0708-6} {\bibfield
  {journal} {\bibinfo  {journal} {Nature materials}\ }\textbf {\bibinfo
  {volume} {19}},\ \bibinfo {pages} {861} (\bibinfo {year} {2020})}\BibitemShut
  {NoStop}%
\bibitem [{\citenamefont {Xu}\ \emph {et~al.}(2020)\citenamefont {Xu},
  \citenamefont {Liu}, \citenamefont {Rhodes}, \citenamefont {Watanabe},
  \citenamefont {Taniguchi}, \citenamefont {Hone}, \citenamefont {Elser},
  \citenamefont {Mak},\ and\ \citenamefont {Shan}}]{xu2020correlated}%
  \BibitemOpen
  \bibfield  {author} {\bibinfo {author} {\bibfnamefont {Y.}~\bibnamefont
  {Xu}}, \bibinfo {author} {\bibfnamefont {S.}~\bibnamefont {Liu}}, \bibinfo
  {author} {\bibfnamefont {D.~A.}\ \bibnamefont {Rhodes}}, \bibinfo {author}
  {\bibfnamefont {K.}~\bibnamefont {Watanabe}}, \bibinfo {author}
  {\bibfnamefont {T.}~\bibnamefont {Taniguchi}}, \bibinfo {author}
  {\bibfnamefont {J.}~\bibnamefont {Hone}}, \bibinfo {author} {\bibfnamefont
  {V.}~\bibnamefont {Elser}}, \bibinfo {author} {\bibfnamefont {K.~F.}\
  \bibnamefont {Mak}},\ and\ \bibinfo {author} {\bibfnamefont {J.}~\bibnamefont
  {Shan}},\ }\href {https://doi.org/10.1038/s41586-020-2868-6} {\bibfield
  {journal} {\bibinfo  {journal} {Nature}\ }\textbf {\bibinfo {volume} {587}},\
  \bibinfo {pages} {214} (\bibinfo {year} {2020})}\BibitemShut {NoStop}%
\bibitem [{\citenamefont {Li}\ \emph {et~al.}(2021)\citenamefont {Li},
  \citenamefont {Jiang}, \citenamefont {Shen}, \citenamefont {Zhang},
  \citenamefont {Li}, \citenamefont {Tao}, \citenamefont {Devakul},
  \citenamefont {Watanabe}, \citenamefont {Taniguchi}, \citenamefont {Fu} \emph
  {et~al.}}]{li2021quantum}%
  \BibitemOpen
  \bibfield  {author} {\bibinfo {author} {\bibfnamefont {T.}~\bibnamefont
  {Li}}, \bibinfo {author} {\bibfnamefont {S.}~\bibnamefont {Jiang}}, \bibinfo
  {author} {\bibfnamefont {B.}~\bibnamefont {Shen}}, \bibinfo {author}
  {\bibfnamefont {Y.}~\bibnamefont {Zhang}}, \bibinfo {author} {\bibfnamefont
  {L.}~\bibnamefont {Li}}, \bibinfo {author} {\bibfnamefont {Z.}~\bibnamefont
  {Tao}}, \bibinfo {author} {\bibfnamefont {T.}~\bibnamefont {Devakul}},
  \bibinfo {author} {\bibfnamefont {K.}~\bibnamefont {Watanabe}}, \bibinfo
  {author} {\bibfnamefont {T.}~\bibnamefont {Taniguchi}}, \bibinfo {author}
  {\bibfnamefont {L.}~\bibnamefont {Fu}}, \emph {et~al.},\ }\href
  {https://doi.org/10.1038/s41586-021-04171-1} {\bibfield  {journal} {\bibinfo
  {journal} {Nature}\ }\textbf {\bibinfo {volume} {600}},\ \bibinfo {pages}
  {641} (\bibinfo {year} {2021})}\BibitemShut {NoStop}%
\bibitem [{\citenamefont {Cai}\ \emph {et~al.}(2023)\citenamefont {Cai},
  \citenamefont {Anderson}, \citenamefont {Wang}, \citenamefont {Zhang},
  \citenamefont {Liu}, \citenamefont {Holtzmann}, \citenamefont {Zhang},
  \citenamefont {Fan}, \citenamefont {Taniguchi}, \citenamefont {Watanabe}
  \emph {et~al.}}]{cai2023signatures}%
  \BibitemOpen
  \bibfield  {author} {\bibinfo {author} {\bibfnamefont {J.}~\bibnamefont
  {Cai}}, \bibinfo {author} {\bibfnamefont {E.}~\bibnamefont {Anderson}},
  \bibinfo {author} {\bibfnamefont {C.}~\bibnamefont {Wang}}, \bibinfo {author}
  {\bibfnamefont {X.}~\bibnamefont {Zhang}}, \bibinfo {author} {\bibfnamefont
  {X.}~\bibnamefont {Liu}}, \bibinfo {author} {\bibfnamefont {W.}~\bibnamefont
  {Holtzmann}}, \bibinfo {author} {\bibfnamefont {Y.}~\bibnamefont {Zhang}},
  \bibinfo {author} {\bibfnamefont {F.}~\bibnamefont {Fan}}, \bibinfo {author}
  {\bibfnamefont {T.}~\bibnamefont {Taniguchi}}, \bibinfo {author}
  {\bibfnamefont {K.}~\bibnamefont {Watanabe}}, \emph {et~al.},\ }\href
  {https://doi.org/10.1038/s41586-023-06289-w} {\bibfield  {journal} {\bibinfo
  {journal} {Nature}\ }\textbf {\bibinfo {volume} {622}},\ \bibinfo {pages}
  {63} (\bibinfo {year} {2023})}\BibitemShut {NoStop}%
\bibitem [{\citenamefont {Zeng}\ \emph {et~al.}(2023)\citenamefont {Zeng},
  \citenamefont {Xia}, \citenamefont {Kang}, \citenamefont {Zhu}, \citenamefont
  {Kn{\"u}ppel}, \citenamefont {Vaswani}, \citenamefont {Watanabe},
  \citenamefont {Taniguchi}, \citenamefont {Mak},\ and\ \citenamefont
  {Shan}}]{zeng2023thermodynamic}%
  \BibitemOpen
  \bibfield  {author} {\bibinfo {author} {\bibfnamefont {Y.}~\bibnamefont
  {Zeng}}, \bibinfo {author} {\bibfnamefont {Z.}~\bibnamefont {Xia}}, \bibinfo
  {author} {\bibfnamefont {K.}~\bibnamefont {Kang}}, \bibinfo {author}
  {\bibfnamefont {J.}~\bibnamefont {Zhu}}, \bibinfo {author} {\bibfnamefont
  {P.}~\bibnamefont {Kn{\"u}ppel}}, \bibinfo {author} {\bibfnamefont
  {C.}~\bibnamefont {Vaswani}}, \bibinfo {author} {\bibfnamefont
  {K.}~\bibnamefont {Watanabe}}, \bibinfo {author} {\bibfnamefont
  {T.}~\bibnamefont {Taniguchi}}, \bibinfo {author} {\bibfnamefont {K.~F.}\
  \bibnamefont {Mak}},\ and\ \bibinfo {author} {\bibfnamefont {J.}~\bibnamefont
  {Shan}},\ }\href {https://doi.org/10.1038/s41586-023-06452-3} {\bibfield
  {journal} {\bibinfo  {journal} {Nature}\ }\textbf {\bibinfo {volume} {622}},\
  \bibinfo {pages} {69} (\bibinfo {year} {2023})}\BibitemShut {NoStop}%
\bibitem [{\citenamefont {Park}\ \emph {et~al.}(2023)\citenamefont {Park},
  \citenamefont {Cai}, \citenamefont {Anderson}, \citenamefont {Zhang},
  \citenamefont {Zhu}, \citenamefont {Liu}, \citenamefont {Wang}, \citenamefont
  {Holtzmann}, \citenamefont {Hu}, \citenamefont {Liu} \emph
  {et~al.}}]{park2023observation}%
  \BibitemOpen
  \bibfield  {author} {\bibinfo {author} {\bibfnamefont {H.}~\bibnamefont
  {Park}}, \bibinfo {author} {\bibfnamefont {J.}~\bibnamefont {Cai}}, \bibinfo
  {author} {\bibfnamefont {E.}~\bibnamefont {Anderson}}, \bibinfo {author}
  {\bibfnamefont {Y.}~\bibnamefont {Zhang}}, \bibinfo {author} {\bibfnamefont
  {J.}~\bibnamefont {Zhu}}, \bibinfo {author} {\bibfnamefont {X.}~\bibnamefont
  {Liu}}, \bibinfo {author} {\bibfnamefont {C.}~\bibnamefont {Wang}}, \bibinfo
  {author} {\bibfnamefont {W.}~\bibnamefont {Holtzmann}}, \bibinfo {author}
  {\bibfnamefont {C.}~\bibnamefont {Hu}}, \bibinfo {author} {\bibfnamefont
  {Z.}~\bibnamefont {Liu}}, \emph {et~al.},\ }\href
  {https://doi.org/10.1038/s41586-023-06536-0} {\bibfield  {journal} {\bibinfo
  {journal} {Nature}\ }\textbf {\bibinfo {volume} {622}},\ \bibinfo {pages}
  {74} (\bibinfo {year} {2023})}\BibitemShut {NoStop}%
\bibitem [{\citenamefont {Xu}\ \emph {et~al.}(2023)\citenamefont {Xu},
  \citenamefont {Sun}, \citenamefont {Jia}, \citenamefont {Liu}, \citenamefont
  {Xu}, \citenamefont {Li}, \citenamefont {Gu}, \citenamefont {Watanabe},
  \citenamefont {Taniguchi}, \citenamefont {Tong}, \citenamefont {Jia},
  \citenamefont {Shi}, \citenamefont {Jiang}, \citenamefont {Zhang},
  \citenamefont {Liu},\ and\ \citenamefont {Li}}]{xu2023observation}%
  \BibitemOpen
  \bibfield  {author} {\bibinfo {author} {\bibfnamefont {F.}~\bibnamefont
  {Xu}}, \bibinfo {author} {\bibfnamefont {Z.}~\bibnamefont {Sun}}, \bibinfo
  {author} {\bibfnamefont {T.}~\bibnamefont {Jia}}, \bibinfo {author}
  {\bibfnamefont {C.}~\bibnamefont {Liu}}, \bibinfo {author} {\bibfnamefont
  {C.}~\bibnamefont {Xu}}, \bibinfo {author} {\bibfnamefont {C.}~\bibnamefont
  {Li}}, \bibinfo {author} {\bibfnamefont {Y.}~\bibnamefont {Gu}}, \bibinfo
  {author} {\bibfnamefont {K.}~\bibnamefont {Watanabe}}, \bibinfo {author}
  {\bibfnamefont {T.}~\bibnamefont {Taniguchi}}, \bibinfo {author}
  {\bibfnamefont {B.}~\bibnamefont {Tong}}, \bibinfo {author} {\bibfnamefont
  {J.}~\bibnamefont {Jia}}, \bibinfo {author} {\bibfnamefont {Z.}~\bibnamefont
  {Shi}}, \bibinfo {author} {\bibfnamefont {S.}~\bibnamefont {Jiang}}, \bibinfo
  {author} {\bibfnamefont {Y.}~\bibnamefont {Zhang}}, \bibinfo {author}
  {\bibfnamefont {X.}~\bibnamefont {Liu}},\ and\ \bibinfo {author}
  {\bibfnamefont {T.}~\bibnamefont {Li}},\ }\href
  {https://doi.org/10.1103/PhysRevX.13.031037} {\bibfield  {journal} {\bibinfo
  {journal} {Phys. Rev. X}\ }\textbf {\bibinfo {volume} {13}},\ \bibinfo
  {pages} {031037} (\bibinfo {year} {2023})}\BibitemShut {NoStop}%
\bibitem [{\citenamefont {Keimer}\ and\ \citenamefont
  {Moore}(2017)}]{Keimer2017}%
  \BibitemOpen
  \bibfield  {author} {\bibinfo {author} {\bibfnamefont {B.}~\bibnamefont
  {Keimer}}\ and\ \bibinfo {author} {\bibfnamefont {J.~E.}\ \bibnamefont
  {Moore}},\ }\href {https://doi.org/10.1038/nphys4302} {\bibfield  {journal}
  {\bibinfo  {journal} {Nature Physics}\ }\textbf {\bibinfo {volume} {13}},\
  \bibinfo {pages} {1045} (\bibinfo {year} {2017})}\BibitemShut {NoStop}%
\bibitem [{\citenamefont {Paschen}\ and\ \citenamefont {Si}(2021)}]{Pas21.1}%
  \BibitemOpen
  \bibfield  {author} {\bibinfo {author} {\bibfnamefont {S.}~\bibnamefont
  {Paschen}}\ and\ \bibinfo {author} {\bibfnamefont {Q.}~\bibnamefont {Si}},\
  }\href {https://doi.org/10.1038/s42254-020-00262-6} {\bibfield  {journal}
  {\bibinfo  {journal} {{Nat.\ Rev.\ Phys.}}\ }\textbf {\bibinfo {volume}
  {3}},\ \bibinfo {pages} {9} (\bibinfo {year} {2021})}\BibitemShut {NoStop}%
\bibitem [{\citenamefont {Andrei}\ and\ \citenamefont
  {MacDonald}(2020)}]{andrei2020graphene}%
  \BibitemOpen
  \bibfield  {author} {\bibinfo {author} {\bibfnamefont {E.~Y.}\ \bibnamefont
  {Andrei}}\ and\ \bibinfo {author} {\bibfnamefont {A.~H.}\ \bibnamefont
  {MacDonald}},\ }\href {https://doi.org/10.1038/s41563-020-00917-w} {\bibfield
   {journal} {\bibinfo  {journal} {Nature materials}\ }\textbf {\bibinfo
  {volume} {19}},\ \bibinfo {pages} {1265} (\bibinfo {year}
  {2020})}\BibitemShut {NoStop}%
\bibitem [{\citenamefont {Xia}\ \emph {et~al.}(2025)\citenamefont {Xia},
  \citenamefont {Han}, \citenamefont {Watanabe}, \citenamefont {Taniguchi},
  \citenamefont {Shan},\ and\ \citenamefont {Mak}}]{Xia2024Unconventional}%
  \BibitemOpen
  \bibfield  {author} {\bibinfo {author} {\bibfnamefont {Y.}~\bibnamefont
  {Xia}}, \bibinfo {author} {\bibfnamefont {Z.}~\bibnamefont {Han}}, \bibinfo
  {author} {\bibfnamefont {K.}~\bibnamefont {Watanabe}}, \bibinfo {author}
  {\bibfnamefont {T.}~\bibnamefont {Taniguchi}}, \bibinfo {author}
  {\bibfnamefont {J.}~\bibnamefont {Shan}},\ and\ \bibinfo {author}
  {\bibfnamefont {K.~F.}\ \bibnamefont {Mak}},\ }\href
  {https://doi.org/10.1038/s41586-024-08116-2} {\bibfield  {journal} {\bibinfo
  {journal} {Nature}\ }\textbf {\bibinfo {volume} {637}},\ \bibinfo {pages}
  {833} (\bibinfo {year} {2025})}\BibitemShut {NoStop}%
\bibitem [{\citenamefont {Guo}\ \emph {et~al.}(2025)\citenamefont {Guo},
  \citenamefont {Pack}, \citenamefont {Swann}, \citenamefont {Holtzman},
  \citenamefont {Cothrine}, \citenamefont {Watanabe}, \citenamefont
  {Taniguchi}, \citenamefont {Mandrus}, \citenamefont {Barmak}, \citenamefont
  {Hone}, \citenamefont {Millis}, \citenamefont {Pasupathy},\ and\
  \citenamefont {Dean}}]{Guo2024Superconductivity}%
  \BibitemOpen
  \bibfield  {author} {\bibinfo {author} {\bibfnamefont {Y.}~\bibnamefont
  {Guo}}, \bibinfo {author} {\bibfnamefont {J.}~\bibnamefont {Pack}}, \bibinfo
  {author} {\bibfnamefont {J.}~\bibnamefont {Swann}}, \bibinfo {author}
  {\bibfnamefont {L.}~\bibnamefont {Holtzman}}, \bibinfo {author}
  {\bibfnamefont {M.}~\bibnamefont {Cothrine}}, \bibinfo {author}
  {\bibfnamefont {K.}~\bibnamefont {Watanabe}}, \bibinfo {author}
  {\bibfnamefont {T.}~\bibnamefont {Taniguchi}}, \bibinfo {author}
  {\bibfnamefont {D.~G.}\ \bibnamefont {Mandrus}}, \bibinfo {author}
  {\bibfnamefont {K.}~\bibnamefont {Barmak}}, \bibinfo {author} {\bibfnamefont
  {J.}~\bibnamefont {Hone}}, \bibinfo {author} {\bibfnamefont {A.~J.}\
  \bibnamefont {Millis}}, \bibinfo {author} {\bibfnamefont {A.}~\bibnamefont
  {Pasupathy}},\ and\ \bibinfo {author} {\bibfnamefont {C.~R.}\ \bibnamefont
  {Dean}},\ }\href {https://doi.org/10.1038/s41586-024-08381-1} {\bibfield
  {journal} {\bibinfo  {journal} {Nature}\ }\textbf {\bibinfo {volume} {637}},\
  \bibinfo {pages} {839} (\bibinfo {year} {2025})}\BibitemShut {NoStop}%
\bibitem [{\citenamefont {Lee}\ \emph {et~al.}(2006)\citenamefont {Lee},
  \citenamefont {Nagaosa},\ and\ \citenamefont {Wen}}]{Lee-RMP06}%
  \BibitemOpen
  \bibfield  {author} {\bibinfo {author} {\bibfnamefont {P.~A.}\ \bibnamefont
  {Lee}}, \bibinfo {author} {\bibfnamefont {N.}~\bibnamefont {Nagaosa}},\ and\
  \bibinfo {author} {\bibfnamefont {X.-G.}\ \bibnamefont {Wen}},\ }\href
  {https://doi.org/10.1103/RevModPhys.78.17} {\bibfield  {journal} {\bibinfo
  {journal} {Rev. Mod. Phys.}\ }\textbf {\bibinfo {volume} {78}},\ \bibinfo
  {pages} {17} (\bibinfo {year} {2006})}\BibitemShut {NoStop}%
\bibitem [{\citenamefont {Si}\ and\ \citenamefont {Steglich}(2010)}]{Si10.2}%
  \BibitemOpen
  \bibfield  {author} {\bibinfo {author} {\bibfnamefont {Q.}~\bibnamefont
  {Si}}\ and\ \bibinfo {author} {\bibfnamefont {F.}~\bibnamefont {Steglich}},\
  }\href {https://doi.org/10.1126/science.1191195} {\bibfield  {journal}
  {\bibinfo  {journal} {{Science}}\ }\textbf {\bibinfo {volume} {{329}}},\
  \bibinfo {pages} {{1161}} (\bibinfo {year} {{2010}})}\BibitemShut {NoStop}%
\bibitem [{\citenamefont {Si}\ \emph {et~al.}(2016)\citenamefont {Si},
  \citenamefont {Yu},\ and\ \citenamefont {Abrahams}}]{Si16.1}%
  \BibitemOpen
  \bibfield  {author} {\bibinfo {author} {\bibfnamefont {Q.}~\bibnamefont
  {Si}}, \bibinfo {author} {\bibfnamefont {R.}~\bibnamefont {Yu}},\ and\
  \bibinfo {author} {\bibfnamefont {E.}~\bibnamefont {Abrahams}},\ }\href
  {https://doi.org/10.1038/natrevmats.2016.17} {\bibfield  {journal} {\bibinfo
  {journal} {Nature Reviews Materials}\ }\textbf {\bibinfo {volume} {1}},\
  \bibinfo {pages} {16017} (\bibinfo {year} {2016})}\BibitemShut {NoStop}%
\bibitem [{\citenamefont {Wu}\ \emph {et~al.}(2018{\natexlab{a}})\citenamefont
  {Wu}, \citenamefont {MacDonald},\ and\ \citenamefont
  {Martin}}]{wu2018theory}%
  \BibitemOpen
  \bibfield  {author} {\bibinfo {author} {\bibfnamefont {F.}~\bibnamefont
  {Wu}}, \bibinfo {author} {\bibfnamefont {A.~H.}\ \bibnamefont {MacDonald}},\
  and\ \bibinfo {author} {\bibfnamefont {I.}~\bibnamefont {Martin}},\ }\href
  {https://doi.org/10.1103/PhysRevLett.121.257001} {\bibfield  {journal}
  {\bibinfo  {journal} {Phys. Rev. Lett.}\ }\textbf {\bibinfo {volume} {121}},\
  \bibinfo {pages} {257001} (\bibinfo {year} {2018}{\natexlab{a}})}\BibitemShut
  {NoStop}%
\bibitem [{\citenamefont {Isobe}\ \emph {et~al.}(2018)\citenamefont {Isobe},
  \citenamefont {Yuan},\ and\ \citenamefont {Fu}}]{isobe2018unconventional}%
  \BibitemOpen
  \bibfield  {author} {\bibinfo {author} {\bibfnamefont {H.}~\bibnamefont
  {Isobe}}, \bibinfo {author} {\bibfnamefont {N.~F.~Q.}\ \bibnamefont {Yuan}},\
  and\ \bibinfo {author} {\bibfnamefont {L.}~\bibnamefont {Fu}},\ }\href
  {https://doi.org/10.1103/PhysRevX.8.041041} {\bibfield  {journal} {\bibinfo
  {journal} {Phys. Rev. X}\ }\textbf {\bibinfo {volume} {8}},\ \bibinfo {pages}
  {041041} (\bibinfo {year} {2018})}\BibitemShut {NoStop}%
\bibitem [{\citenamefont {Guo}\ \emph {et~al.}(2018)\citenamefont {Guo},
  \citenamefont {Zhu}, \citenamefont {Feng},\ and\ \citenamefont
  {Scalettar}}]{guo2018pairing}%
  \BibitemOpen
  \bibfield  {author} {\bibinfo {author} {\bibfnamefont {H.}~\bibnamefont
  {Guo}}, \bibinfo {author} {\bibfnamefont {X.}~\bibnamefont {Zhu}}, \bibinfo
  {author} {\bibfnamefont {S.}~\bibnamefont {Feng}},\ and\ \bibinfo {author}
  {\bibfnamefont {R.~T.}\ \bibnamefont {Scalettar}},\ }\href
  {https://doi.org/10.1103/PhysRevB.97.235453} {\bibfield  {journal} {\bibinfo
  {journal} {Phys. Rev. B}\ }\textbf {\bibinfo {volume} {97}},\ \bibinfo
  {pages} {235453} (\bibinfo {year} {2018})}\BibitemShut {NoStop}%
\bibitem [{\citenamefont {Xu}\ and\ \citenamefont
  {Balents}(2018)}]{xu2018topological}%
  \BibitemOpen
  \bibfield  {author} {\bibinfo {author} {\bibfnamefont {C.}~\bibnamefont
  {Xu}}\ and\ \bibinfo {author} {\bibfnamefont {L.}~\bibnamefont {Balents}},\
  }\href {https://doi.org/10.1103/PhysRevLett.121.087001} {\bibfield  {journal}
  {\bibinfo  {journal} {Phys. Rev. Lett.}\ }\textbf {\bibinfo {volume} {121}},\
  \bibinfo {pages} {087001} (\bibinfo {year} {2018})}\BibitemShut {NoStop}%
\bibitem [{\citenamefont {Kim}\ \emph {et~al.}(2025)\citenamefont {Kim},
  \citenamefont {Mendez-Valderrama}, \citenamefont {Wang},\ and\ \citenamefont
  {Chowdhury}}]{Debanjan2024}%
  \BibitemOpen
  \bibfield  {author} {\bibinfo {author} {\bibfnamefont {S.}~\bibnamefont
  {Kim}}, \bibinfo {author} {\bibfnamefont {J.~F.}\ \bibnamefont
  {Mendez-Valderrama}}, \bibinfo {author} {\bibfnamefont {X.}~\bibnamefont
  {Wang}},\ and\ \bibinfo {author} {\bibfnamefont {D.}~\bibnamefont
  {Chowdhury}},\ }\href {https://doi.org/10.1038/s41467-025-56816-8} {\bibfield
   {journal} {\bibinfo  {journal} {Nature Communications}\ }\textbf {\bibinfo
  {volume} {16}},\ \bibinfo {pages} {1701} (\bibinfo {year}
  {2025})}\BibitemShut {NoStop}%
\bibitem [{\citenamefont {Schrade}\ and\ \citenamefont
  {Fu}(2024)}]{Schrade2021Nematic}%
  \BibitemOpen
  \bibfield  {author} {\bibinfo {author} {\bibfnamefont {C.}~\bibnamefont
  {Schrade}}\ and\ \bibinfo {author} {\bibfnamefont {L.}~\bibnamefont {Fu}},\
  }\href {https://doi.org/10.1103/PhysRevB.110.035143} {\bibfield  {journal}
  {\bibinfo  {journal} {Phys. Rev. B}\ }\textbf {\bibinfo {volume} {110}},\
  \bibinfo {pages} {035143} (\bibinfo {year} {2024})}\BibitemShut {NoStop}%
\bibitem [{\citenamefont {Chen}\ and\ \citenamefont
  {Sheng}(2023)}]{Chen2023Singlet}%
  \BibitemOpen
  \bibfield  {author} {\bibinfo {author} {\bibfnamefont {F.}~\bibnamefont
  {Chen}}\ and\ \bibinfo {author} {\bibfnamefont {D.~N.}\ \bibnamefont
  {Sheng}},\ }\href {https://doi.org/10.1103/PhysRevB.108.L201110} {\bibfield
  {journal} {\bibinfo  {journal} {Phys. Rev. B}\ }\textbf {\bibinfo {volume}
  {108}},\ \bibinfo {pages} {L201110} (\bibinfo {year} {2023})}\BibitemShut
  {NoStop}%
\bibitem [{\citenamefont {Zhou}\ and\ \citenamefont
  {Zhang}(2023)}]{Zhou2023Chiral}%
  \BibitemOpen
  \bibfield  {author} {\bibinfo {author} {\bibfnamefont {B.}~\bibnamefont
  {Zhou}}\ and\ \bibinfo {author} {\bibfnamefont {Y.-H.}\ \bibnamefont
  {Zhang}},\ }\href {https://doi.org/10.1103/PhysRevB.108.155111} {\bibfield
  {journal} {\bibinfo  {journal} {Phys. Rev. B}\ }\textbf {\bibinfo {volume}
  {108}},\ \bibinfo {pages} {155111} (\bibinfo {year} {2023})}\BibitemShut
  {NoStop}%
\bibitem [{\citenamefont {{Christos}}\ \emph {et~al.}(2024)\citenamefont
  {{Christos}}, \citenamefont {{Bonetti}},\ and\ \citenamefont
  {{Scheurer}}}]{Christos2024Approximate}%
  \BibitemOpen
  \bibfield  {author} {\bibinfo {author} {\bibfnamefont {M.}~\bibnamefont
  {{Christos}}}, \bibinfo {author} {\bibfnamefont {P.~M.}\ \bibnamefont
  {{Bonetti}}},\ and\ \bibinfo {author} {\bibfnamefont {M.~S.}\ \bibnamefont
  {{Scheurer}}},\ }\href {https://doi.org/10.48550/arXiv.2407.02393} {\bibfield
   {journal} {\bibinfo  {journal} {arXiv e-prints}\ ,\ \bibinfo {eid}
  {arXiv:2407.02393}} (\bibinfo {year} {2024})},\ \Eprint
  {https://arxiv.org/abs/2407.02393} {arXiv:2407.02393 [cond-mat.supr-con]}
  \BibitemShut {NoStop}%
\bibitem [{\citenamefont {Wu}\ \emph {et~al.}(2019)\citenamefont {Wu},
  \citenamefont {Lovorn}, \citenamefont {Tutuc}, \citenamefont {Martin},\ and\
  \citenamefont {MacDonald}}]{Wu2019topological}%
  \BibitemOpen
  \bibfield  {author} {\bibinfo {author} {\bibfnamefont {F.}~\bibnamefont
  {Wu}}, \bibinfo {author} {\bibfnamefont {T.}~\bibnamefont {Lovorn}}, \bibinfo
  {author} {\bibfnamefont {E.}~\bibnamefont {Tutuc}}, \bibinfo {author}
  {\bibfnamefont {I.}~\bibnamefont {Martin}},\ and\ \bibinfo {author}
  {\bibfnamefont {A.~H.}\ \bibnamefont {MacDonald}},\ }\href
  {https://doi.org/10.1103/PhysRevLett.122.086402} {\bibfield  {journal}
  {\bibinfo  {journal} {Phys. Rev. Lett.}\ }\textbf {\bibinfo {volume} {122}},\
  \bibinfo {pages} {086402} (\bibinfo {year} {2019})}\BibitemShut {NoStop}%
\bibitem [{\citenamefont {Devakul}\ \emph {et~al.}(2021)\citenamefont
  {Devakul}, \citenamefont {Cr{\'e}pel}, \citenamefont {Zhang},\ and\
  \citenamefont {Fu}}]{Devakul2021Magic}%
  \BibitemOpen
  \bibfield  {author} {\bibinfo {author} {\bibfnamefont {T.}~\bibnamefont
  {Devakul}}, \bibinfo {author} {\bibfnamefont {V.}~\bibnamefont {Cr{\'e}pel}},
  \bibinfo {author} {\bibfnamefont {Y.}~\bibnamefont {Zhang}},\ and\ \bibinfo
  {author} {\bibfnamefont {L.}~\bibnamefont {Fu}},\ }\href
  {https://doi.org/10.1038/s41467-021-27042-9} {\bibfield  {journal} {\bibinfo
  {journal} {Nature Communications}\ }\textbf {\bibinfo {volume} {12}},\
  \bibinfo {pages} {6730} (\bibinfo {year} {2021})}\BibitemShut {NoStop}%
\bibitem [{\citenamefont {Reddy}\ \emph {et~al.}(2023)\citenamefont {Reddy},
  \citenamefont {Alsallom}, \citenamefont {Zhang}, \citenamefont {Devakul},\
  and\ \citenamefont {Fu}}]{Reddy2023Fractional}%
  \BibitemOpen
  \bibfield  {author} {\bibinfo {author} {\bibfnamefont {A.~P.}\ \bibnamefont
  {Reddy}}, \bibinfo {author} {\bibfnamefont {F.}~\bibnamefont {Alsallom}},
  \bibinfo {author} {\bibfnamefont {Y.}~\bibnamefont {Zhang}}, \bibinfo
  {author} {\bibfnamefont {T.}~\bibnamefont {Devakul}},\ and\ \bibinfo {author}
  {\bibfnamefont {L.}~\bibnamefont {Fu}},\ }\href
  {https://doi.org/10.1103/PhysRevB.108.085117} {\bibfield  {journal} {\bibinfo
   {journal} {Phys. Rev. B}\ }\textbf {\bibinfo {volume} {108}},\ \bibinfo
  {pages} {085117} (\bibinfo {year} {2023})}\BibitemShut {NoStop}%
\bibitem [{sup()}]{supplemental_material}%
  \BibitemOpen
  \href@noop {} {}\bibinfo {note} {See the Supplemental Material, which
  includes \cite{Yu2024Fractional, bradley2010mathematical, Pizzi2020wannier,
  Laturia2018, Yu2012U1slave}, for additional discussion about the model
  construction, numerical simulation and interacting phase
  diagram.}\BibitemShut {Stop}%
\bibitem [{\citenamefont {Hu}\ and\ \citenamefont {Si}(2023)}]{hu2023coupled}%
  \BibitemOpen
  \bibfield  {author} {\bibinfo {author} {\bibfnamefont {H.}~\bibnamefont
  {Hu}}\ and\ \bibinfo {author} {\bibfnamefont {Q.}~\bibnamefont {Si}},\ }\href
  {https://doi.org/10.1126/sciadv.abm0100} {\bibfield  {journal} {\bibinfo
  {journal} {Science Advances}\ }\textbf {\bibinfo {volume} {9}},\ \bibinfo
  {pages} {eadg0028} (\bibinfo {year} {2023})}\BibitemShut {NoStop}%
\bibitem [{\citenamefont {Chen}\ \emph {et~al.}(2024)\citenamefont {Chen},
  \citenamefont {Xie}, \citenamefont {Sur}, \citenamefont {Hu}, \citenamefont
  {Paschen}, \citenamefont {Cano},\ and\ \citenamefont
  {Si}}]{chen2024emergent}%
  \BibitemOpen
  \bibfield  {author} {\bibinfo {author} {\bibfnamefont {L.}~\bibnamefont
  {Chen}}, \bibinfo {author} {\bibfnamefont {F.}~\bibnamefont {Xie}}, \bibinfo
  {author} {\bibfnamefont {S.}~\bibnamefont {Sur}}, \bibinfo {author}
  {\bibfnamefont {H.}~\bibnamefont {Hu}}, \bibinfo {author} {\bibfnamefont
  {S.}~\bibnamefont {Paschen}}, \bibinfo {author} {\bibfnamefont
  {J.}~\bibnamefont {Cano}},\ and\ \bibinfo {author} {\bibfnamefont
  {Q.}~\bibnamefont {Si}},\ }\href {https://doi.org/10.5281/zenodo.11247849}
  {\bibfield  {journal} {\bibinfo  {journal} {Nature Communications}\ }\textbf
  {\bibinfo {volume} {15}},\ \bibinfo {pages} {5242} (\bibinfo {year}
  {2024})}\BibitemShut {NoStop}%
\bibitem [{\citenamefont {{Chen}}\ \emph {et~al.}(2023)\citenamefont {{Chen}},
  \citenamefont {{Xie}}, \citenamefont {{Sur}}, \citenamefont {{Hu}},
  \citenamefont {{Paschen}}, \citenamefont {{Cano}},\ and\ \citenamefont
  {{Si}}}]{chen2023metallic}%
  \BibitemOpen
  \bibfield  {author} {\bibinfo {author} {\bibfnamefont {L.}~\bibnamefont
  {{Chen}}}, \bibinfo {author} {\bibfnamefont {F.}~\bibnamefont {{Xie}}},
  \bibinfo {author} {\bibfnamefont {S.}~\bibnamefont {{Sur}}}, \bibinfo
  {author} {\bibfnamefont {H.}~\bibnamefont {{Hu}}}, \bibinfo {author}
  {\bibfnamefont {S.}~\bibnamefont {{Paschen}}}, \bibinfo {author}
  {\bibfnamefont {J.}~\bibnamefont {{Cano}}},\ and\ \bibinfo {author}
  {\bibfnamefont {Q.}~\bibnamefont {{Si}}},\ }\href
  {https://doi.org/10.48550/arXiv.2307.09431} {\bibfield  {journal} {\bibinfo
  {journal} {arXiv e-prints}\ ,\ \bibinfo {eid} {arXiv:2307.09431}} (\bibinfo
  {year} {2023})},\ \Eprint {https://arxiv.org/abs/2307.09431}
  {arXiv:2307.09431 [cond-mat.str-el]} \BibitemShut {NoStop}%
\bibitem [{\citenamefont {Lai}\ \emph {et~al.}(2018)\citenamefont {Lai},
  \citenamefont {Grefe}, \citenamefont {Paschen},\ and\ \citenamefont
  {Si}}]{lai2018weyl}%
  \BibitemOpen
  \bibfield  {author} {\bibinfo {author} {\bibfnamefont {H.-H.}\ \bibnamefont
  {Lai}}, \bibinfo {author} {\bibfnamefont {S.~E.}\ \bibnamefont {Grefe}},
  \bibinfo {author} {\bibfnamefont {S.}~\bibnamefont {Paschen}},\ and\ \bibinfo
  {author} {\bibfnamefont {Q.}~\bibnamefont {Si}},\ }\href
  {https://doi.org/10.1073/pnas.1715851115} {\bibfield  {journal} {\bibinfo
  {journal} {Proceedings of the National Academy of Sciences}\ }\textbf
  {\bibinfo {volume} {115}},\ \bibinfo {pages} {93} (\bibinfo {year}
  {2018})}\BibitemShut {NoStop}%
\bibitem [{\citenamefont {Dzsaber}\ \emph {et~al.}(2021)\citenamefont
  {Dzsaber}, \citenamefont {Yan}, \citenamefont {Taupin}, \citenamefont
  {Eguchi}, \citenamefont {Prokofiev}, \citenamefont {Shiroka}, \citenamefont
  {Blaha}, \citenamefont {Rubel}, \citenamefont {Grefe}, \citenamefont {Lai}
  \emph {et~al.}}]{dzsaber2021giant}%
  \BibitemOpen
  \bibfield  {author} {\bibinfo {author} {\bibfnamefont {S.}~\bibnamefont
  {Dzsaber}}, \bibinfo {author} {\bibfnamefont {X.}~\bibnamefont {Yan}},
  \bibinfo {author} {\bibfnamefont {M.}~\bibnamefont {Taupin}}, \bibinfo
  {author} {\bibfnamefont {G.}~\bibnamefont {Eguchi}}, \bibinfo {author}
  {\bibfnamefont {A.}~\bibnamefont {Prokofiev}}, \bibinfo {author}
  {\bibfnamefont {T.}~\bibnamefont {Shiroka}}, \bibinfo {author} {\bibfnamefont
  {P.}~\bibnamefont {Blaha}}, \bibinfo {author} {\bibfnamefont
  {O.}~\bibnamefont {Rubel}}, \bibinfo {author} {\bibfnamefont {S.~E.}\
  \bibnamefont {Grefe}}, \bibinfo {author} {\bibfnamefont {H.-H.}\ \bibnamefont
  {Lai}}, \emph {et~al.},\ }\href {https://doi.org/10.1073/pnas.2013386118}
  {\bibfield  {journal} {\bibinfo  {journal} {Proceedings of the National
  Academy of Sciences}\ }\textbf {\bibinfo {volume} {118}},\ \bibinfo {pages}
  {e2013386118} (\bibinfo {year} {2021})}\BibitemShut {NoStop}%
\bibitem [{\citenamefont {Hu}\ \emph {et~al.}(2024)\citenamefont {Hu},
  \citenamefont {Chen},\ and\ \citenamefont {Si}}]{Hu-Natphys2024}%
  \BibitemOpen
  \bibfield  {author} {\bibinfo {author} {\bibfnamefont {H.}~\bibnamefont
  {Hu}}, \bibinfo {author} {\bibfnamefont {L.}~\bibnamefont {Chen}},\ and\
  \bibinfo {author} {\bibfnamefont {Q.}~\bibnamefont {Si}},\ }\href
  {https://doi.org/10.1038/s41567-024-02679-7} {\bibfield  {journal} {\bibinfo
  {journal} {Nat. Phys.}\ }\textbf {\bibinfo {volume} {20}},\ \bibinfo {pages}
  {1863} (\bibinfo {year} {2024})}\BibitemShut {NoStop}%
\bibitem [{\citenamefont {{Hu}}\ \emph {et~al.}(2022)\citenamefont {{Hu}},
  \citenamefont {{Chen}},\ and\ \citenamefont {{Si}}}]{Hu2022extended}%
  \BibitemOpen
  \bibfield  {author} {\bibinfo {author} {\bibfnamefont {H.}~\bibnamefont
  {{Hu}}}, \bibinfo {author} {\bibfnamefont {L.}~\bibnamefont {{Chen}}},\ and\
  \bibinfo {author} {\bibfnamefont {Q.}~\bibnamefont {{Si}}},\ }\href
  {https://doi.org/10.48550/arXiv.2210.14197} {\bibfield  {journal} {\bibinfo
  {journal} {arXiv e-prints}\ ,\ \bibinfo {eid} {arXiv:2210.14197}} (\bibinfo
  {year} {2022})},\ \Eprint {https://arxiv.org/abs/2210.14197}
  {arXiv:2210.14197 [cond-mat.str-el]} \BibitemShut {NoStop}%
\bibitem [{\citenamefont {{Hu}}\ \emph {et~al.}(2021)\citenamefont {{Hu}},
  \citenamefont {{Cai}}, \citenamefont {{Chen}}, \citenamefont {{Deng}},
  \citenamefont {{Pixley}}, \citenamefont {{Ingersent}},\ and\ \citenamefont
  {{Si}}}]{Hu2021-sc}%
  \BibitemOpen
  \bibfield  {author} {\bibinfo {author} {\bibfnamefont {H.}~\bibnamefont
  {{Hu}}}, \bibinfo {author} {\bibfnamefont {A.}~\bibnamefont {{Cai}}},
  \bibinfo {author} {\bibfnamefont {L.}~\bibnamefont {{Chen}}}, \bibinfo
  {author} {\bibfnamefont {L.}~\bibnamefont {{Deng}}}, \bibinfo {author}
  {\bibfnamefont {J.~H.}\ \bibnamefont {{Pixley}}}, \bibinfo {author}
  {\bibfnamefont {K.}~\bibnamefont {{Ingersent}}},\ and\ \bibinfo {author}
  {\bibfnamefont {Q.}~\bibnamefont {{Si}}},\ }\href
  {https://doi.org/10.48550/arXiv.2109.13224} {\bibfield  {journal} {\bibinfo
  {journal} {arXiv e-prints}\ ,\ \bibinfo {eid} {arXiv:2109.13224}} (\bibinfo
  {year} {2021})},\ \Eprint {https://arxiv.org/abs/2109.13224}
  {arXiv:2109.13224 [cond-mat.str-el]} \BibitemShut {NoStop}%
\bibitem [{\citenamefont {Liu}\ \emph {et~al.}(2013)\citenamefont {Liu},
  \citenamefont {Shan}, \citenamefont {Yao}, \citenamefont {Yao},\ and\
  \citenamefont {Xiao}}]{Liu2013Three}%
  \BibitemOpen
  \bibfield  {author} {\bibinfo {author} {\bibfnamefont {G.-B.}\ \bibnamefont
  {Liu}}, \bibinfo {author} {\bibfnamefont {W.-Y.}\ \bibnamefont {Shan}},
  \bibinfo {author} {\bibfnamefont {Y.}~\bibnamefont {Yao}}, \bibinfo {author}
  {\bibfnamefont {W.}~\bibnamefont {Yao}},\ and\ \bibinfo {author}
  {\bibfnamefont {D.}~\bibnamefont {Xiao}},\ }\href
  {https://doi.org/10.1103/PhysRevB.88.085433} {\bibfield  {journal} {\bibinfo
  {journal} {Phys. Rev. B}\ }\textbf {\bibinfo {volume} {88}},\ \bibinfo
  {pages} {085433} (\bibinfo {year} {2013})}\BibitemShut {NoStop}%
\bibitem [{\citenamefont {Kormányos}\ \emph {et~al.}(2015)\citenamefont
  {Kormányos}, \citenamefont {Burkard}, \citenamefont {Gmitra}, \citenamefont
  {Fabian}, \citenamefont {Zólyomi}, \citenamefont {Drummond},\ and\
  \citenamefont {Fal’ko}}]{Kormanyos2015kp}%
  \BibitemOpen
  \bibfield  {author} {\bibinfo {author} {\bibfnamefont {A.}~\bibnamefont
  {Kormányos}}, \bibinfo {author} {\bibfnamefont {G.}~\bibnamefont {Burkard}},
  \bibinfo {author} {\bibfnamefont {M.}~\bibnamefont {Gmitra}}, \bibinfo
  {author} {\bibfnamefont {J.}~\bibnamefont {Fabian}}, \bibinfo {author}
  {\bibfnamefont {V.}~\bibnamefont {Zólyomi}}, \bibinfo {author}
  {\bibfnamefont {N.~D.}\ \bibnamefont {Drummond}},\ and\ \bibinfo {author}
  {\bibfnamefont {V.}~\bibnamefont {Fal’ko}},\ }\href
  {https://doi.org/10.1088/2053-1583/2/2/022001} {\bibfield  {journal}
  {\bibinfo  {journal} {2D Materials}\ }\textbf {\bibinfo {volume} {2}},\
  \bibinfo {pages} {022001} (\bibinfo {year} {2015})}\BibitemShut {NoStop}%
\bibitem [{\citenamefont {Wu}\ \emph {et~al.}(2018{\natexlab{b}})\citenamefont
  {Wu}, \citenamefont {Lovorn}, \citenamefont {Tutuc},\ and\ \citenamefont
  {MacDonald}}]{Wu2018Hubbard}%
  \BibitemOpen
  \bibfield  {author} {\bibinfo {author} {\bibfnamefont {F.}~\bibnamefont
  {Wu}}, \bibinfo {author} {\bibfnamefont {T.}~\bibnamefont {Lovorn}}, \bibinfo
  {author} {\bibfnamefont {E.}~\bibnamefont {Tutuc}},\ and\ \bibinfo {author}
  {\bibfnamefont {A.~H.}\ \bibnamefont {MacDonald}},\ }\href
  {https://doi.org/10.1103/PhysRevLett.121.026402} {\bibfield  {journal}
  {\bibinfo  {journal} {Phys. Rev. Lett.}\ }\textbf {\bibinfo {volume} {121}},\
  \bibinfo {pages} {026402} (\bibinfo {year} {2018}{\natexlab{b}})}\BibitemShut
  {NoStop}%
\bibitem [{\citenamefont {Fazekas}(1999)}]{fazekas1999lecture}%
  \BibitemOpen
  \bibfield  {author} {\bibinfo {author} {\bibfnamefont {P.}~\bibnamefont
  {Fazekas}},\ }\href@noop {} {\emph {\bibinfo {title} {Lecture notes on
  electron correlation and magnetism}}},\ Vol.~\bibinfo {volume} {5}\ (\bibinfo
   {publisher} {World scientific},\ \bibinfo {year} {1999})\BibitemShut
  {NoStop}%
\bibitem [{\citenamefont {Kotliar}\ and\ \citenamefont
  {Ruckenstein}(1986)}]{Kotliar1986}%
  \BibitemOpen
  \bibfield  {author} {\bibinfo {author} {\bibfnamefont {G.}~\bibnamefont
  {Kotliar}}\ and\ \bibinfo {author} {\bibfnamefont {A.~E.}\ \bibnamefont
  {Ruckenstein}},\ }\href {https://doi.org/10.1103/PhysRevLett.57.1362}
  {\bibfield  {journal} {\bibinfo  {journal} {Phys. Rev. Lett.}\ }\textbf
  {\bibinfo {volume} {57}},\ \bibinfo {pages} {1362} (\bibinfo {year}
  {1986})}\BibitemShut {NoStop}%
\bibitem [{\citenamefont {Hassan}\ and\ \citenamefont {de'
  Medici}(2010)}]{Hassan2010slave}%
  \BibitemOpen
  \bibfield  {author} {\bibinfo {author} {\bibfnamefont {S.~R.}\ \bibnamefont
  {Hassan}}\ and\ \bibinfo {author} {\bibfnamefont {L.}~\bibnamefont {de'
  Medici}},\ }\href {https://doi.org/10.1103/PhysRevB.81.035106} {\bibfield
  {journal} {\bibinfo  {journal} {Phys. Rev. B}\ }\textbf {\bibinfo {volume}
  {81}},\ \bibinfo {pages} {035106} (\bibinfo {year} {2010})}\BibitemShut
  {NoStop}%
\bibitem [{\citenamefont {Pixley}\ \emph {et~al.}(2015)\citenamefont {Pixley},
  \citenamefont {Cai},\ and\ \citenamefont {Si}}]{Pixley-c-edmft2015}%
  \BibitemOpen
  \bibfield  {author} {\bibinfo {author} {\bibfnamefont {J.~H.}\ \bibnamefont
  {Pixley}}, \bibinfo {author} {\bibfnamefont {A.}~\bibnamefont {Cai}},\ and\
  \bibinfo {author} {\bibfnamefont {Q.}~\bibnamefont {Si}},\ }\href
  {https://doi.org/10.1103/PhysRevB.91.125127} {\bibfield  {journal} {\bibinfo
  {journal} {Phys. Rev. B}\ }\textbf {\bibinfo {volume} {91}},\ \bibinfo
  {pages} {125127} (\bibinfo {year} {2015})}\BibitemShut {NoStop}%
\bibitem [{\citenamefont {Si}\ \emph {et~al.}(2001)\citenamefont {Si},
  \citenamefont {Rabello}, \citenamefont {Ingersent},\ and\ \citenamefont
  {Smith}}]{Si01.1}%
  \BibitemOpen
  \bibfield  {author} {\bibinfo {author} {\bibfnamefont {Q.}~\bibnamefont
  {Si}}, \bibinfo {author} {\bibfnamefont {S.}~\bibnamefont {Rabello}},
  \bibinfo {author} {\bibfnamefont {K.}~\bibnamefont {Ingersent}},\ and\
  \bibinfo {author} {\bibfnamefont {J.}~\bibnamefont {Smith}},\ }\href
  {https://doi.org/10.1038/35101507} {\bibfield  {journal} {\bibinfo  {journal}
  {Nature}\ }\textbf {\bibinfo {volume} {413}},\ \bibinfo {pages} {804}
  (\bibinfo {year} {2001})}\BibitemShut {NoStop}%
\bibitem [{\citenamefont {Mielke}(1992)}]{Mielke_1992}%
  \BibitemOpen
  \bibfield  {author} {\bibinfo {author} {\bibfnamefont {A.}~\bibnamefont
  {Mielke}},\ }\href {https://doi.org/10.1088/0305-4470/25/16/011} {\bibfield
  {journal} {\bibinfo  {journal} {Journal of Physics A: Mathematical and
  General}\ }\textbf {\bibinfo {volume} {25}},\ \bibinfo {pages} {4335}
  (\bibinfo {year} {1992})}\BibitemShut {NoStop}%
\bibitem [{\citenamefont {Lin}\ \emph {et~al.}(2024)\citenamefont {Lin},
  \citenamefont {Liu},\ and\ \citenamefont {Moore}}]{Lin2023Complex}%
  \BibitemOpen
  \bibfield  {author} {\bibinfo {author} {\bibfnamefont {Y.-P.}\ \bibnamefont
  {Lin}}, \bibinfo {author} {\bibfnamefont {C.}~\bibnamefont {Liu}},\ and\
  \bibinfo {author} {\bibfnamefont {J.~E.}\ \bibnamefont {Moore}},\ }\href
  {https://doi.org/10.1103/PhysRevB.110.L041121} {\bibfield  {journal}
  {\bibinfo  {journal} {Phys. Rev. B}\ }\textbf {\bibinfo {volume} {110}},\
  \bibinfo {pages} {L041121} (\bibinfo {year} {2024})}\BibitemShut {NoStop}%
\bibitem [{\citenamefont {Kirchner}\ \emph {et~al.}(2020)\citenamefont
  {Kirchner}, \citenamefont {Paschen}, \citenamefont {Chen}, \citenamefont
  {Wirth}, \citenamefont {Feng}, \citenamefont {Thompson},\ and\ \citenamefont
  {Si}}]{Kirchner_RMP}%
  \BibitemOpen
  \bibfield  {author} {\bibinfo {author} {\bibfnamefont {S.}~\bibnamefont
  {Kirchner}}, \bibinfo {author} {\bibfnamefont {S.}~\bibnamefont {Paschen}},
  \bibinfo {author} {\bibfnamefont {Q.}~\bibnamefont {Chen}}, \bibinfo {author}
  {\bibfnamefont {S.}~\bibnamefont {Wirth}}, \bibinfo {author} {\bibfnamefont
  {D.}~\bibnamefont {Feng}}, \bibinfo {author} {\bibfnamefont {J.~D.}\
  \bibnamefont {Thompson}},\ and\ \bibinfo {author} {\bibfnamefont
  {Q.}~\bibnamefont {Si}},\ }\href
  {https://doi.org/10.1103/RevModPhys.92.011002} {\bibfield  {journal}
  {\bibinfo  {journal} {Rev. Mod. Phys.}\ }\textbf {\bibinfo {volume} {92}},\
  \bibinfo {pages} {011002} (\bibinfo {year} {2020})}\BibitemShut {NoStop}%
\bibitem [{\citenamefont {Paschen}\ \emph {et~al.}(2004)\citenamefont
  {Paschen}, \citenamefont {L\"uhmann}, \citenamefont {Wirth}, \citenamefont
  {Gegenwart}, \citenamefont {Trovarelli}, \citenamefont {Geibel},
  \citenamefont {Steglich}, \citenamefont {Coleman},\ and\ \citenamefont
  {Si}}]{Pas04.1}%
  \BibitemOpen
  \bibfield  {author} {\bibinfo {author} {\bibfnamefont {S.}~\bibnamefont
  {Paschen}}, \bibinfo {author} {\bibfnamefont {T.}~\bibnamefont {L\"uhmann}},
  \bibinfo {author} {\bibfnamefont {S.}~\bibnamefont {Wirth}}, \bibinfo
  {author} {\bibfnamefont {P.}~\bibnamefont {Gegenwart}}, \bibinfo {author}
  {\bibfnamefont {O.}~\bibnamefont {Trovarelli}}, \bibinfo {author}
  {\bibfnamefont {C.}~\bibnamefont {Geibel}}, \bibinfo {author} {\bibfnamefont
  {F.}~\bibnamefont {Steglich}}, \bibinfo {author} {\bibfnamefont
  {P.}~\bibnamefont {Coleman}},\ and\ \bibinfo {author} {\bibfnamefont
  {Q.}~\bibnamefont {Si}},\ }\href {https://doi.org/10.1038/nature03129}
  {\bibfield  {journal} {\bibinfo  {journal} {{Nature}}\ }\textbf {\bibinfo
  {volume} {432}},\ \bibinfo {pages} {881} (\bibinfo {year}
  {2004})}\BibitemShut {NoStop}%
\bibitem [{\citenamefont {Zhao}\ \emph {et~al.}(2023)\citenamefont {Zhao},
  \citenamefont {Shen}, \citenamefont {Tao}, \citenamefont {Han}, \citenamefont
  {Kang}, \citenamefont {Watanabe}, \citenamefont {Taniguchi}, \citenamefont
  {Mak},\ and\ \citenamefont {Shan}}]{zhao2022gate}%
  \BibitemOpen
  \bibfield  {author} {\bibinfo {author} {\bibfnamefont {W.}~\bibnamefont
  {Zhao}}, \bibinfo {author} {\bibfnamefont {B.}~\bibnamefont {Shen}}, \bibinfo
  {author} {\bibfnamefont {Z.}~\bibnamefont {Tao}}, \bibinfo {author}
  {\bibfnamefont {Z.}~\bibnamefont {Han}}, \bibinfo {author} {\bibfnamefont
  {K.}~\bibnamefont {Kang}}, \bibinfo {author} {\bibfnamefont {K.}~\bibnamefont
  {Watanabe}}, \bibinfo {author} {\bibfnamefont {T.}~\bibnamefont {Taniguchi}},
  \bibinfo {author} {\bibfnamefont {K.~F.}\ \bibnamefont {Mak}},\ and\ \bibinfo
  {author} {\bibfnamefont {J.}~\bibnamefont {Shan}},\ }\href
  {https://doi.org/10.1038/s41586-023-05800-7} {\bibfield  {journal} {\bibinfo
  {journal} {Nature}\ }\textbf {\bibinfo {volume} {616}},\ \bibinfo {pages}
  {61} (\bibinfo {year} {2023})}\BibitemShut {NoStop}%
\bibitem [{\citenamefont {Zhao}\ \emph {et~al.}(2024)\citenamefont {Zhao},
  \citenamefont {Shen}, \citenamefont {Tao}, \citenamefont {Kim}, \citenamefont
  {Kn{\"u}ppel}, \citenamefont {Han}, \citenamefont {Zhang}, \citenamefont
  {Watanabe}, \citenamefont {Taniguchi}, \citenamefont {Chowdhury},
  \citenamefont {Shan},\ and\ \citenamefont {Mak}}]{zhao2023emergence}%
  \BibitemOpen
  \bibfield  {author} {\bibinfo {author} {\bibfnamefont {W.}~\bibnamefont
  {Zhao}}, \bibinfo {author} {\bibfnamefont {B.}~\bibnamefont {Shen}}, \bibinfo
  {author} {\bibfnamefont {Z.}~\bibnamefont {Tao}}, \bibinfo {author}
  {\bibfnamefont {S.}~\bibnamefont {Kim}}, \bibinfo {author} {\bibfnamefont
  {P.}~\bibnamefont {Kn{\"u}ppel}}, \bibinfo {author} {\bibfnamefont
  {Z.}~\bibnamefont {Han}}, \bibinfo {author} {\bibfnamefont {Y.}~\bibnamefont
  {Zhang}}, \bibinfo {author} {\bibfnamefont {K.}~\bibnamefont {Watanabe}},
  \bibinfo {author} {\bibfnamefont {T.}~\bibnamefont {Taniguchi}}, \bibinfo
  {author} {\bibfnamefont {D.}~\bibnamefont {Chowdhury}}, \bibinfo {author}
  {\bibfnamefont {J.}~\bibnamefont {Shan}},\ and\ \bibinfo {author}
  {\bibfnamefont {K.~F.}\ \bibnamefont {Mak}},\ }\href
  {https://doi.org/10.1038/s41567-024-02636-4} {\bibfield  {journal} {\bibinfo
  {journal} {Nature Physics}\ }\textbf {\bibinfo {volume} {20}},\ \bibinfo
  {pages} {1772} (\bibinfo {year} {2024})}\BibitemShut {NoStop}%
\bibitem [{\citenamefont {Guerci}\ \emph {et~al.}(2023)\citenamefont {Guerci},
  \citenamefont {Wang}, \citenamefont {Zang}, \citenamefont {Cano},
  \citenamefont {Pixley},\ and\ \citenamefont {Millis}}]{Guerci2023Chiral}%
  \BibitemOpen
  \bibfield  {author} {\bibinfo {author} {\bibfnamefont {D.}~\bibnamefont
  {Guerci}}, \bibinfo {author} {\bibfnamefont {J.}~\bibnamefont {Wang}},
  \bibinfo {author} {\bibfnamefont {J.}~\bibnamefont {Zang}}, \bibinfo {author}
  {\bibfnamefont {J.}~\bibnamefont {Cano}}, \bibinfo {author} {\bibfnamefont
  {J.~H.}\ \bibnamefont {Pixley}},\ and\ \bibinfo {author} {\bibfnamefont
  {A.}~\bibnamefont {Millis}},\ }\href {https://doi.org/10.1126/sciadv.ade7701}
  {\bibfield  {journal} {\bibinfo  {journal} {Science Advances}\ }\textbf
  {\bibinfo {volume} {9}},\ \bibinfo {pages} {eade7701} (\bibinfo {year}
  {2023})}\BibitemShut {NoStop}%
\bibitem [{\citenamefont {Xie}\ \emph {et~al.}(2024)\citenamefont {Xie},
  \citenamefont {Chen},\ and\ \citenamefont {Si}}]{Xie2024Kondo}%
  \BibitemOpen
  \bibfield  {author} {\bibinfo {author} {\bibfnamefont {F.}~\bibnamefont
  {Xie}}, \bibinfo {author} {\bibfnamefont {L.}~\bibnamefont {Chen}},\ and\
  \bibinfo {author} {\bibfnamefont {Q.}~\bibnamefont {Si}},\ }\href
  {https://doi.org/10.1103/PhysRevResearch.6.013219} {\bibfield  {journal}
  {\bibinfo  {journal} {Phys. Rev. Res.}\ }\textbf {\bibinfo {volume} {6}},\
  \bibinfo {pages} {013219} (\bibinfo {year} {2024})}\BibitemShut {NoStop}%
\bibitem [{\citenamefont {Ramires}\ and\ \citenamefont {Lado}(2021)}]{Ram2021}%
  \BibitemOpen
  \bibfield  {author} {\bibinfo {author} {\bibfnamefont {A.}~\bibnamefont
  {Ramires}}\ and\ \bibinfo {author} {\bibfnamefont {J.~L.}\ \bibnamefont
  {Lado}},\ }\href {https://doi.org/10.1103/PhysRevLett.127.026401} {\bibfield
  {journal} {\bibinfo  {journal} {Phys. Rev. Lett.}\ }\textbf {\bibinfo
  {volume} {127}},\ \bibinfo {pages} {026401} (\bibinfo {year}
  {2021})}\BibitemShut {NoStop}%
\bibitem [{\citenamefont {Song}\ and\ \citenamefont
  {Bernevig}(2022)}]{Song2022}%
  \BibitemOpen
  \bibfield  {author} {\bibinfo {author} {\bibfnamefont {Z.-D.}\ \bibnamefont
  {Song}}\ and\ \bibinfo {author} {\bibfnamefont {B.~A.}\ \bibnamefont
  {Bernevig}},\ }\href {https://doi.org/10.1103/PhysRevLett.129.047601}
  {\bibfield  {journal} {\bibinfo  {journal} {Phys. Rev. Lett.}\ }\textbf
  {\bibinfo {volume} {129}},\ \bibinfo {pages} {047601} (\bibinfo {year}
  {2022})}\BibitemShut {NoStop}%
\bibitem [{\citenamefont {Kumar}\ \emph {et~al.}(2022)\citenamefont {Kumar},
  \citenamefont {Hu}, \citenamefont {MacDonald},\ and\ \citenamefont
  {Potter}}]{Kumar2022}%
  \BibitemOpen
  \bibfield  {author} {\bibinfo {author} {\bibfnamefont {A.}~\bibnamefont
  {Kumar}}, \bibinfo {author} {\bibfnamefont {N.~C.}\ \bibnamefont {Hu}},
  \bibinfo {author} {\bibfnamefont {A.~H.}\ \bibnamefont {MacDonald}},\ and\
  \bibinfo {author} {\bibfnamefont {A.~C.}\ \bibnamefont {Potter}},\ }\href
  {https://doi.org/10.1103/PhysRevB.106.L041116} {\bibfield  {journal}
  {\bibinfo  {journal} {Phys. Rev. B}\ }\textbf {\bibinfo {volume} {106}},\
  \bibinfo {pages} {L041116} (\bibinfo {year} {2022})}\BibitemShut {NoStop}%
\bibitem [{\citenamefont {Datta}\ \emph {et~al.}(2023)\citenamefont {Datta},
  \citenamefont {Calder{\'o}n}, \citenamefont {Camjayi},\ and\ \citenamefont
  {Bascones}}]{Datta2023Heavy}%
  \BibitemOpen
  \bibfield  {author} {\bibinfo {author} {\bibfnamefont {A.}~\bibnamefont
  {Datta}}, \bibinfo {author} {\bibfnamefont {M.~J.}\ \bibnamefont
  {Calder{\'o}n}}, \bibinfo {author} {\bibfnamefont {A.}~\bibnamefont
  {Camjayi}},\ and\ \bibinfo {author} {\bibfnamefont {E.}~\bibnamefont
  {Bascones}},\ }\href {https://doi.org/10.1038/s41467-023-40754-4} {\bibfield
  {journal} {\bibinfo  {journal} {Nature Communications}\ }\textbf {\bibinfo
  {volume} {14}},\ \bibinfo {pages} {5036} (\bibinfo {year}
  {2023})}\BibitemShut {NoStop}%
\bibitem [{\citenamefont {Huang}\ \emph {et~al.}(2024)\citenamefont {Huang},
  \citenamefont {Chen}, \citenamefont {Huang}, \citenamefont {Setty},
  \citenamefont {Gao}, \citenamefont {Shi}, \citenamefont {Liu}, \citenamefont
  {Zhang}, \citenamefont {Yilmaz}, \citenamefont {Vescovo}, \citenamefont
  {Hashimoto}, \citenamefont {Lu}, \citenamefont {Yakobson}, \citenamefont
  {Dai}, \citenamefont {Chu}, \citenamefont {Si},\ and\ \citenamefont
  {Yi}}]{Huang2023np}%
  \BibitemOpen
  \bibfield  {author} {\bibinfo {author} {\bibfnamefont {J.}~\bibnamefont
  {Huang}}, \bibinfo {author} {\bibfnamefont {L.}~\bibnamefont {Chen}},
  \bibinfo {author} {\bibfnamefont {Y.}~\bibnamefont {Huang}}, \bibinfo
  {author} {\bibfnamefont {C.}~\bibnamefont {Setty}}, \bibinfo {author}
  {\bibfnamefont {B.}~\bibnamefont {Gao}}, \bibinfo {author} {\bibfnamefont
  {Y.}~\bibnamefont {Shi}}, \bibinfo {author} {\bibfnamefont {Z.}~\bibnamefont
  {Liu}}, \bibinfo {author} {\bibfnamefont {Y.}~\bibnamefont {Zhang}}, \bibinfo
  {author} {\bibfnamefont {T.}~\bibnamefont {Yilmaz}}, \bibinfo {author}
  {\bibfnamefont {E.}~\bibnamefont {Vescovo}}, \bibinfo {author} {\bibfnamefont
  {M.}~\bibnamefont {Hashimoto}}, \bibinfo {author} {\bibfnamefont
  {D.}~\bibnamefont {Lu}}, \bibinfo {author} {\bibfnamefont {B.~I.}\
  \bibnamefont {Yakobson}}, \bibinfo {author} {\bibfnamefont {P.}~\bibnamefont
  {Dai}}, \bibinfo {author} {\bibfnamefont {J.-H.}\ \bibnamefont {Chu}},
  \bibinfo {author} {\bibfnamefont {Q.}~\bibnamefont {Si}},\ and\ \bibinfo
  {author} {\bibfnamefont {M.}~\bibnamefont {Yi}},\ }\href
  {https://doi.org/10.1038/s41567-023-02362-3} {\bibfield  {journal} {\bibinfo
  {journal} {Nat. Phys.}\ }\textbf {\bibinfo {volume} {20}},\ \bibinfo {pages}
  {603} (\bibinfo {year} {2024})}\BibitemShut {NoStop}%
\bibitem [{\citenamefont {Yu}\ \emph {et~al.}(2024)\citenamefont {Yu},
  \citenamefont {Herzog-Arbeitman}, \citenamefont {Wang}, \citenamefont
  {Vafek}, \citenamefont {Bernevig},\ and\ \citenamefont
  {Regnault}}]{Yu2024Fractional}%
  \BibitemOpen
  \bibfield  {author} {\bibinfo {author} {\bibfnamefont {J.}~\bibnamefont
  {Yu}}, \bibinfo {author} {\bibfnamefont {J.}~\bibnamefont
  {Herzog-Arbeitman}}, \bibinfo {author} {\bibfnamefont {M.}~\bibnamefont
  {Wang}}, \bibinfo {author} {\bibfnamefont {O.}~\bibnamefont {Vafek}},
  \bibinfo {author} {\bibfnamefont {B.~A.}\ \bibnamefont {Bernevig}},\ and\
  \bibinfo {author} {\bibfnamefont {N.}~\bibnamefont {Regnault}},\ }\href
  {https://doi.org/10.1103/PhysRevB.109.045147} {\bibfield  {journal} {\bibinfo
   {journal} {Phys. Rev. B}\ }\textbf {\bibinfo {volume} {109}},\ \bibinfo
  {pages} {045147} (\bibinfo {year} {2024})}\BibitemShut {NoStop}%
\bibitem [{\citenamefont {Bradley}\ and\ \citenamefont
  {Cracknell}(2010)}]{bradley2010mathematical}%
  \BibitemOpen
  \bibfield  {author} {\bibinfo {author} {\bibfnamefont {C.}~\bibnamefont
  {Bradley}}\ and\ \bibinfo {author} {\bibfnamefont {A.}~\bibnamefont
  {Cracknell}},\ }\href@noop {} {\emph {\bibinfo {title} {The mathematical
  theory of symmetry in solids: representation theory for point groups and
  space groups}}}\ (\bibinfo  {publisher} {Oxford University Press},\ \bibinfo
  {year} {2010})\BibitemShut {NoStop}%
\bibitem [{\citenamefont {Pizzi}\ \emph {et~al.}(2020)\citenamefont {Pizzi},
  \citenamefont {Vitale}, \citenamefont {Arita}, \citenamefont {Blügel},
  \citenamefont {Freimuth}, \citenamefont {G{\'{e}}ranton}, \citenamefont
  {Gibertini}, \citenamefont {Gresch}, \citenamefont {Johnson}, \citenamefont
  {Koretsune}, \citenamefont {Iba{\~{n}}ez-Azpiroz}, \citenamefont {Lee},
  \citenamefont {Lihm}, \citenamefont {Marchand}, \citenamefont {Marrazzo},
  \citenamefont {Mokrousov}, \citenamefont {Mustafa}, \citenamefont {Nohara},
  \citenamefont {Nomura}, \citenamefont {Paulatto}, \citenamefont
  {Ponc{\'{e}}}, \citenamefont {Ponweiser}, \citenamefont {Qiao}, \citenamefont
  {Thöle}, \citenamefont {Tsirkin}, \citenamefont {Wierzbowska}, \citenamefont
  {Marzari}, \citenamefont {Vanderbilt}, \citenamefont {Souza}, \citenamefont
  {Mostofi},\ and\ \citenamefont {Yates}}]{Pizzi2020wannier}%
  \BibitemOpen
  \bibfield  {author} {\bibinfo {author} {\bibfnamefont {G.}~\bibnamefont
  {Pizzi}}, \bibinfo {author} {\bibfnamefont {V.}~\bibnamefont {Vitale}},
  \bibinfo {author} {\bibfnamefont {R.}~\bibnamefont {Arita}}, \bibinfo
  {author} {\bibfnamefont {S.}~\bibnamefont {Blügel}}, \bibinfo {author}
  {\bibfnamefont {F.}~\bibnamefont {Freimuth}}, \bibinfo {author}
  {\bibfnamefont {G.}~\bibnamefont {G{\'{e}}ranton}}, \bibinfo {author}
  {\bibfnamefont {M.}~\bibnamefont {Gibertini}}, \bibinfo {author}
  {\bibfnamefont {D.}~\bibnamefont {Gresch}}, \bibinfo {author} {\bibfnamefont
  {C.}~\bibnamefont {Johnson}}, \bibinfo {author} {\bibfnamefont
  {T.}~\bibnamefont {Koretsune}}, \bibinfo {author} {\bibfnamefont
  {J.}~\bibnamefont {Iba{\~{n}}ez-Azpiroz}}, \bibinfo {author} {\bibfnamefont
  {H.}~\bibnamefont {Lee}}, \bibinfo {author} {\bibfnamefont {J.-M.}\
  \bibnamefont {Lihm}}, \bibinfo {author} {\bibfnamefont {D.}~\bibnamefont
  {Marchand}}, \bibinfo {author} {\bibfnamefont {A.}~\bibnamefont {Marrazzo}},
  \bibinfo {author} {\bibfnamefont {Y.}~\bibnamefont {Mokrousov}}, \bibinfo
  {author} {\bibfnamefont {J.~I.}\ \bibnamefont {Mustafa}}, \bibinfo {author}
  {\bibfnamefont {Y.}~\bibnamefont {Nohara}}, \bibinfo {author} {\bibfnamefont
  {Y.}~\bibnamefont {Nomura}}, \bibinfo {author} {\bibfnamefont
  {L.}~\bibnamefont {Paulatto}}, \bibinfo {author} {\bibfnamefont
  {S.}~\bibnamefont {Ponc{\'{e}}}}, \bibinfo {author} {\bibfnamefont
  {T.}~\bibnamefont {Ponweiser}}, \bibinfo {author} {\bibfnamefont
  {J.}~\bibnamefont {Qiao}}, \bibinfo {author} {\bibfnamefont {F.}~\bibnamefont
  {Thöle}}, \bibinfo {author} {\bibfnamefont {S.~S.}\ \bibnamefont {Tsirkin}},
  \bibinfo {author} {\bibfnamefont {M.}~\bibnamefont {Wierzbowska}}, \bibinfo
  {author} {\bibfnamefont {N.}~\bibnamefont {Marzari}}, \bibinfo {author}
  {\bibfnamefont {D.}~\bibnamefont {Vanderbilt}}, \bibinfo {author}
  {\bibfnamefont {I.}~\bibnamefont {Souza}}, \bibinfo {author} {\bibfnamefont
  {A.~A.}\ \bibnamefont {Mostofi}},\ and\ \bibinfo {author} {\bibfnamefont
  {J.~R.}\ \bibnamefont {Yates}},\ }\href
  {https://doi.org/10.1088/1361-648x/ab51ff} {\bibfield  {journal} {\bibinfo
  {journal} {Journal of Physics: Condensed Matter}\ }\textbf {\bibinfo {volume}
  {32}},\ \bibinfo {pages} {165902} (\bibinfo {year} {2020})}\BibitemShut
  {NoStop}%
\bibitem [{\citenamefont {Laturia}\ \emph {et~al.}(2018)\citenamefont
  {Laturia}, \citenamefont {Van~de Put},\ and\ \citenamefont
  {Vandenberghe}}]{Laturia2018}%
  \BibitemOpen
  \bibfield  {author} {\bibinfo {author} {\bibfnamefont {A.}~\bibnamefont
  {Laturia}}, \bibinfo {author} {\bibfnamefont {M.~L.}\ \bibnamefont {Van~de
  Put}},\ and\ \bibinfo {author} {\bibfnamefont {W.~G.}\ \bibnamefont
  {Vandenberghe}},\ }\href {https://doi.org/10.1038/s41699-018-0050-x}
  {\bibfield  {journal} {\bibinfo  {journal} {npj 2D Materials and
  Applications}\ }\textbf {\bibinfo {volume} {2}},\ \bibinfo {pages} {6}
  (\bibinfo {year} {2018})}\BibitemShut {NoStop}%
\bibitem [{\citenamefont {Yu}\ and\ \citenamefont {Si}(2012)}]{Yu2012U1slave}%
  \BibitemOpen
  \bibfield  {author} {\bibinfo {author} {\bibfnamefont {R.}~\bibnamefont
  {Yu}}\ and\ \bibinfo {author} {\bibfnamefont {Q.}~\bibnamefont {Si}},\ }\href
  {https://doi.org/10.1103/PhysRevB.86.085104} {\bibfield  {journal} {\bibinfo
  {journal} {Phys. Rev. B}\ }\textbf {\bibinfo {volume} {86}},\ \bibinfo
  {pages} {085104} (\bibinfo {year} {2012})}\BibitemShut {NoStop}%
\end{thebibliography}%
\bibliographystyle{apsrev4-2}

\setcounter{secnumdepth}{3}
\clearpage

\onecolumngrid
\beginsupplement

\section*{Supplemental Materials}

\section{Continuum model}\label{sec:app-continuum}

In this section, we provide a brief discussion about the continuum model of twisted homo-bilayer TMDC, which was introduced in Ref.~\cite{Wu2019topological}.
With the inter-layer and intra-layer moir\'e potential considered, the low-energy effective Hamiltonian of twisted bilayer WSe$_2$ can be written as:
\begin{equation}\label{eqn:tmd-hamiltonian}
    H = \left(
        \begin{array}{cc}
            \frac{\nabla^2}{2m^*} + v_+(\mathbf{r}) + \frac{\varepsilon_D}{2} & t(\mathbf{r})\\
            t^*(\mathbf{r}) & \frac{\nabla^2}{2m^*} + v_-(\mathbf{r}) - \frac{\varepsilon_D}{2}
        \end{array}
    \right)\,.
\end{equation}
in which $m^*$ is the effective mass of the hole pocket, $v_\pm(\mathbf{r})$ is the moir\'e potential in the top and bottom layers, respectively, and $t(\mathbf{r})$ is the inter-layer tunneling term. 
The displacement field potential $\varepsilon_D$ is also considered in the Hamiltonian.
In the absence of the displacement field, the Hamiltonian Eq.~(\ref{eqn:tmd-hamiltonian}) possesses both $C_{3z}$ and $C_{2y}T$ symmetries, while the displacement field term will break the $C_{2y}T$ symmetry.
Under the lowest-harmonic approximation \cite{Wu2019topological}, the $C_{3z}$ symmetry enforces the intra-layer periodic potential to have the following form:
{\begin{equation}
    v_{\ell}(\mathbf{r}) = 2\tilde{v} \sum_{j=1,3,5} \cos(\mathbf{g}_j\cdot\mathbf{r} + \ell \psi)\,,~~\ell = \pm\,,
\end{equation}
in which the reciprocal lattice vectors are defined as $\mathbf{g}_1 = -\mathbf{g}_4 = \mathbf{b}_1$, $\mathbf{g}_3 = -\mathbf{g}_6 = \mathbf{b}_2 - \mathbf{b}_1$ and $\mathbf{g}_5 = -\mathbf{g}_2 = -\mathbf{b}_2$.
Similarly, the inter-layer tunneling term can be written as:
\begin{equation}
    t(\mathbf{r}) = w \sum_{j = 1,2,3}e^{i\vq_j \cdot \mathbf{r}}\,,    
\end{equation}
in which the definition of the three $\vq_j$ vectors can be found in Fig.~1(b) in the main text.
By fitting the continuum Hamiltonian Eq.~(\ref{eqn:tmd-hamiltonian}) with {\it ab initio} simulations, the parameters for twisted bilayer $\rm WSe_2$ are determined to be $a_0 = 3.317\,\text{\AA}$, $m^* = 0.43m_e$, $\tilde{v} = 9\,{\rm meV}$, $\psi = 128^\circ$ and $w = 18\,\rm meV$, in which $m_e$ is the mass of the free electron \cite{Devakul2021Magic}.

This continuum Hamiltonian can be numerically diagonalized using the plane wave basis, whose eigenstate wave function (component in layer $\ell = \pm 1$) can be represented in the following form:
\begin{equation}\label{eqn:bloch-eigenstate}
    \psi_{ n\vk}(\mathbf{r}, \ell) = \frac{1}{\sqrt{\Omega_{\rm tot}}}\sum_{\mathbf{Q} \in \mathcal{Q}_\ell} u_{\mathbf{Q}, n}(\vk)e^{i(\vk - \mathbf{Q})\cdot \mathbf{r}}\,.
\end{equation}
Here $\Omega_{\rm tot}$ is the volume of the sample, and the momentum space grid $\mathcal{Q}_{\ell} $ is defined as $\mathcal{Q}_0 + \ell \vq_1$, in which $\mathcal{Q}_0$ stands for moir\'e reciprocal lattice.
Band structure plots shown in Fig.~2 in the main text are computed from this continuum model.

\section{Band topology}\label{sec:app-bands}

In this section, we will discuss the band structure and the associated band topology of twisted bilayer WSe$_2$ with different moir\'e potential strengths.
As indicated by the continuum model Hamiltonian in Eq.~(\ref{eqn:tmd-hamiltonian}), the band structure is mostly determined by the twisting angle $\theta$ and the moir\'e potential strength $\tilde{v}$ and $w$, and is expected to be sensitive to such parameters~\cite{Yu2024Fractional}.

Here, we study the band structure and the topology of the top bands in the case of tWSe$_2$ with the angle $\psi = 128^\circ$ and effective mass $m^* = 0.43m_e$ fixed.
In Fig.~\ref{fig:gaps}, we show the direct band gaps between the first and the second top-most bands $\Delta_{12}$, and between the second and the third top-most bands $\Delta_{23}$ as functions of the moir\'e potential strength $\tilde{v}$ and $w$, with the twisting angles $\theta = 3.65^\circ$ and $\theta = 1.43^\circ$.
The moir\'e potential parameters used in Ref.~\cite{Devakul2021Magic} is labeled by the star symbol, and the Chern numbers of the top two bands are also labeled in blue text in Fig.~\ref{fig:gaps}.

In the case with $\theta = 3.65^\circ$ [see Figs.~\ref{fig:gaps}(a-b)], we vary the strength of the intra-layer potential $\tilde{v}$ from $5$ to $15\,\rm meV$, and the inter-layer potential $w$ from $10$ to $30\,\rm meV$. 
With a suitable choice of 
the moir\'e potential strength, it is possible to obtain a band structure with the top two bands carrying zero total Chern number at twisting angle $\theta = 3.65^\circ$.
In particular, when the intra-layer and the inter-layer potentials are modified to, for example, $(\tilde{v}, w) = (13{\rm\, meV}, 15{\rm\, meV})$, the top two bands will also carry opposite Chern numbers at twisting angle $\theta = 3.65^\circ$.
Due to the reduced size of the moir\'e unit cell, the typical single band width and band gap are around $20\sim 30\,\rm meV$, while the projected interaction is around $30\sim 85\,\rm meV$, as we will discuss in Sec.~\ref{app:interaction-strength}. 
Thus, the regime of our interest is pertinent to this twist angle.
Similarly, when the twisting angle $\theta$ is increased to $\sim 5^\circ$, the top two bands will also have opposite valley Chern numbers when the moir\'e potential strengths are, for example, $(\tilde{v}, w) = (13\,{\rm meV}, 15\,{\rm meV})$. 
The band widths of these active bands will also be further increased to around $80\,\rm meV$, with an on-site interaction strength $U \approx 35\sim 100 \,\rm meV$.
The low-energy effective theory is again in the regime of our interest, and as such we still expect that the same mechanism can drive a superconducting phase around the quantum critical point.

In the parameter space of Figs.~\ref{fig:gaps}(a-b), the parameter variation does not strongly affect the energetics and the single-particle wave function of the top-most moir\'e band and the band top of the second moir\'e band, which are the active degrees of freedom in tWSe$_2$. 
Therefore, our parameter choice still effectively represents the key features of the low-energy physics in tWSe$_2$ at a twist angle of $\theta = 3.65^\circ$.
In the following numerical calculations in Secs.~\ref{sec:app-tb}, \ref{app:interaction-strength} and \ref{sec:app-numerical}, we will adopt the moir\'e potential parameters specified in the previous paragraph.

\begin{figure}[t]
    \centering
    \includegraphics[width=0.5\linewidth]{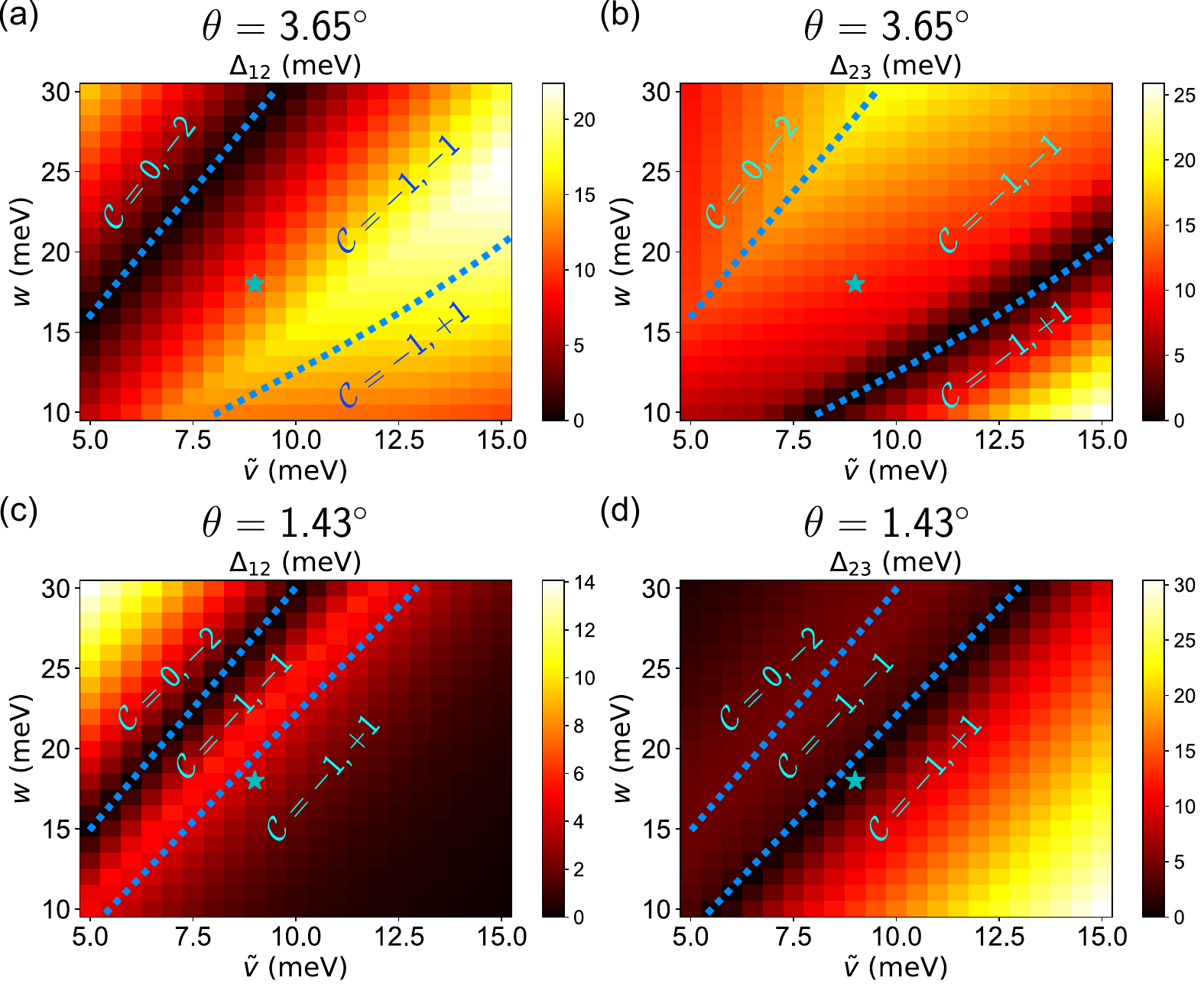}
    \caption{(a-b) The smallest direct band gap between the first and second top-most bands ($\Delta_{12}$), and between the second and the third top-most bands ($\Delta_{23}$) as functions of the moir\'e potential strength $\tilde{v}$ and $w$, with twisting angle $\theta = 3.65^\circ$. 
    (c-d) The smallest direct band gaps with twisting angle $\theta = 1.43^\circ$.
    The parameters used in Ref.~\cite{Devakul2021Magic} are labeled by the cyan star.
    The dashed lines indicate topological phase transitions, with the Chern numbers of the top two bands labeled.
    Here we set the displacement field potential strength to be $\varepsilon_D = 0\,\rm meV$, and $\psi = 128^\circ$.
    }
    \label{fig:gaps}
\end{figure}

\section{Tight-binding parameters}\label{sec:app-tb}

In the band structure shown in Fig.~1 of the main text, the top two bands carry opposite Chern numbers, which allow us to construct two $C_{3z}$ symmetric moir\'e orbitals.
In both cases, the top-most band has $C_{3z}$ eigenvalues of $\omega = e^{i2\pi/3}$, $\omega^*$ and $\omega^*$ at $\gamma$, $\kappa$ and $\kappa'$ points, respectively, while the second band has $C_{3z}$ eigenvalues of $\omega, 1, 1$ at these three points.
These symmetry eigenvalues are compatible with the summation of two $^2E$ orbitals at the $1b$ [$\bm{\tau}_1 = (2\mathbf{a}_1 + \mathbf{a}_2)/3$] and $1c$ [$\bm{\tau}_2 = (\mathbf{a}_1 + 2\mathbf{a}_2)/3$] Wyckoff positions of the space group $P3$ \cite{bradley2010mathematical}.
In the absence of a displacement field, the two orbitals are related by $C_{2y} T$ symmetry.

Using \textsc{Wannier90} \cite{Pizzi2020wannier} program, we are able to obtain a two-orbital tight-binding model, which faithfully describe the two active bands of the continuum model in a single valley.
The effective single valley two-orbital tight binding model has the following form:
\begin{equation}
    H^K_{\rm TB} = \sum_{\mathbf{R}_0\mathbf{R}\alpha\beta}  t_{\alpha\beta}(\mathbf{R}) c^\dagger_{\mathbf{R} + \mathbf{R}_0,\alpha,\uparrow} c_{\mathbf{R}_0,\beta,\uparrow}\,.
\end{equation}
Here $\alpha, \beta \in \{1,2\}$ indicate the two orbitals, and $\mathbf{R},\mathbf{R}_0$ stand for Bravais lattice vectors.
Since the two valleys (spins) are related by time-reversal symmetry, the tight-binding parameters for the $K'$ valley are complex conjugate of those for the $K$ valley:
\begin{equation}
    H^{K'}_{\rm TB} = \sum_{\mathbf{R}_0\mathbf{R}\alpha\beta} t^*_{\alpha\beta}(\mathbf{R}) c^\dagger_{\mathbf{R} + \mathbf{R}_0,\alpha,\downarrow} c_{\mathbf{R}_0,\beta,\downarrow}\,.
\end{equation}

The value of $\epsilon_f$ mentioned in the main text is estimated by the on-site energy.
The average hybridization strength $V_{\rm hyb}$ is estimated by the following expression:
\begin{equation}
    V_{\rm hyb} = \sqrt{\frac{1}{\Omega_{\rm BZ}}\int d^2k \Big{|}\sum_{\mathbf{R}}t_{12}(\mathbf{R}) e^{-i\vk\cdot (\mathbf{R} + \bm{\tau}_1 - \bm{\tau}_2)}\Big{|}^2}\,,
\end{equation}
in which $t_{12}(\mathbf{R})$ stands for the hoppings between orbital 1 and orbital 2 in this tight-binding model.

The values of the hopping parameters for the two-orbital tight-binding model, which are greater than $0.1\,\rm meV$, with parameters $(\tilde{v}, w, \varepsilon_D, \theta) = (9 \,{\rm meV}, 18\,{\rm meV}, 3\,{\rm meV}, 1.43^\circ)$, which are the ones used in the main text, are listed in Table~\ref{tab:parameters}.
In addition, the tight-binding model parameters with model parameters $(\tilde{v}, w, \varepsilon_D, \theta) = (13 \,{\rm meV}, 15\,{\rm meV}, 15\,{\rm meV}, 3.65^\circ)$ are also listed in Table \ref{tab:parameters2}.

\begin{table}
    \begin{center}
    \begin{tabular}{c|c|c|c|c|c|c}
        \hline
        \# & $\alpha$ & $\beta$ & $R_1$ & $R_2$ & $|t_{\alpha\beta}(\mathbf{R})|~(\rm meV)$ & ${\rm arg}~t_{\alpha\beta}(\mathbf{R})$ \\
        \hline $1$ & $1$ & $1$ & $0$ & $0$ & 27.60 & $0.0^\circ$ \\ 
        \hline $2$ & $2$ & $2$ & $0$ & $0$ & 25.63 & $0.0^\circ$ \\ 
        \hline $3$ & $2$ & $1$ & $0$ & $-1$ & 1.39 & $36.5^\circ$ \\ 
        \hline $4$ & $1$ & $2$ & $0$ & $1$ & 1.39 & $-36.5^\circ$ \\ 
        \hline $5$ & $1$ & $2$ & $-1$ & $0$ & 1.39 & $-36.5^\circ$ \\ 
        \hline $6$ & $2$ & $1$ & $0$ & $0$ & 1.39 & $36.5^\circ$ \\ 
        \hline $7$ & $1$ & $2$ & $0$ & $0$ & 1.39 & $-36.5^\circ$ \\ 
        \hline $8$ & $2$ & $1$ & $1$ & $0$ & 1.39 & $36.5^\circ$ \\ 
        \hline $9$ & $2$ & $2$ & $-1$ & $-1$ & 0.52 & $-121.9^\circ$ \\ 
        \hline $10$ & $2$ & $2$ & $-1$ & $0$ & 0.52 & $121.9^\circ$ \\ 
        \hline $11$ & $2$ & $2$ & $0$ & $-1$ & 0.52 & $121.9^\circ$ \\ 
        \hline $12$ & $2$ & $2$ & $0$ & $1$ & 0.52 & $-121.9^\circ$ \\ 
        \hline $13$ & $2$ & $2$ & $1$ & $0$ & 0.52 & $-121.9^\circ$ \\ 
        \hline $14$ & $2$ & $2$ & $1$ & $1$ & 0.52 & $121.9^\circ$ \\ 
        \hline $15$ & $1$ & $1$ & $-1$ & $-1$ & 0.43 & $125.3^\circ$ \\ 
        \hline $16$ & $1$ & $1$ & $-1$ & $0$ & 0.43 & $-125.3^\circ$ \\ 
        \hline $17$ & $1$ & $1$ & $0$ & $-1$ & 0.43 & $-125.3^\circ$ \\ 
        \hline $18$ & $1$ & $1$ & $0$ & $1$ & 0.43 & $125.3^\circ$ \\ 
        \hline $19$ & $1$ & $1$ & $1$ & $0$ & 0.43 & $125.3^\circ$ \\ 
        \hline $20$ & $1$ & $1$ & $1$ & $1$ & 0.43 & $-125.3^\circ$ \\ 
        \hline $21$ & $2$ & $1$ & $-1$ & $-1$ & 0.16 & $-143.5^\circ$ \\ 
        \hline $22$ & $1$ & $2$ & $-1$ & $-1$ & 0.16 & $143.5^\circ$ \\ 
        \hline $23$ & $1$ & $2$ & $-1$ & $1$ & 0.16 & $143.5^\circ$ \\ 
        \hline $24$ & $2$ & $1$ & $1$ & $-1$ & 0.16 & $-143.5^\circ$ \\ 
        \hline $25$ & $2$ & $1$ & $1$ & $1$ & 0.16 & $-143.5^\circ$ \\ 
        \hline $26$ & $1$ & $2$ & $1$ & $1$ & 0.16 & $143.5^\circ$ \\ 
        \hline $27$ & $1$ & $2$ & $-2$ & $0$ & 0.12 & $-38.1^\circ$ \\ 
        \hline $28$ & $2$ & $1$ & $-1$ & $-2$ & 0.12 & $38.1^\circ$ \\ 
        \hline $29$ & $1$ & $2$ & $0$ & $-1$ & 0.12 & $-38.1^\circ$ \\ 
        \hline $30$ & $2$ & $1$ & $0$ & $1$ & 0.12 & $38.1^\circ$ \\ 
        \hline $31$ & $1$ & $2$ & $1$ & $2$ & 0.12 & $-38.1^\circ$ \\ 
        \hline $32$ & $2$ & $1$ & $2$ & $0$ & 0.12 & $38.1^\circ$ \\ 
        \hline $33$ & $1$ & $2$ & $-2$ & $-1$ & 0.12 & $-34.9^\circ$ \\ 
        \hline $34$ & $2$ & $1$ & $-1$ & $0$ & 0.12 & $34.9^\circ$ \\ 
        \hline $35$ & $2$ & $1$ & $0$ & $-2$ & 0.12 & $34.9^\circ$ \\ 
        \hline $36$ & $1$ & $2$ & $0$ & $2$ & 0.12 & $-34.9^\circ$ \\ 
        \hline $37$ & $1$ & $2$ & $1$ & $0$ & 0.12 & $-34.9^\circ$ \\ 
        \hline $38$ & $2$ & $1$ & $2$ & $1$ & 0.12 & $34.9^\circ$ \\
        \hline
    \end{tabular}
    \end{center}
    \caption{
        Parameters of the single valley tight-binding model with $\theta = 1.43^\circ$ and $\varepsilon_D = 3\,\rm meV$. The variables $R_1,R_2\in\mathbb{Z}$ stand for the components of the hopping vector $ \mathbf{R} = R_1 \mathbf{a}_1 + R_2 \mathbf{a}_2$.
        Note that the first two rows are the on-site energies of the two orbitals.
    }
    \label{tab:parameters}
\end{table}

\begin{table}
    \begin{center}
    \begin{tabular}{c|c|c|c|c|c|c}
        \hline
        \# & $\alpha$ & $\beta$ & $R_1$ & $R_2$ & $|t_{\alpha\beta}(\mathbf{R})|~(\rm meV)$ & ${\rm arg}~t_{\alpha\beta}(\mathbf{R})$ \\
        \hline $1$ & $2$ & $2$ & $0$ & $0$ & 13.76 & $180.0^\circ$ \\ 
        \hline $2$ & $2$ & $2$ & $-1$ & $-1$ & 4.38 & $-115.3^\circ$ \\ 
        \hline $3$ & $2$ & $2$ & $-1$ & $0$ & 4.38 & $115.3^\circ$ \\ 
        \hline $4$ & $2$ & $2$ & $0$ & $-1$ & 4.38 & $115.3^\circ$ \\ 
        \hline $5$ & $2$ & $2$ & $0$ & $1$ & 4.38 & $-115.3^\circ$ \\ 
        \hline $6$ & $2$ & $2$ & $1$ & $0$ & 4.38 & $-115.3^\circ$ \\ 
        \hline $7$ & $2$ & $2$ & $1$ & $1$ & 4.38 & $115.3^\circ$ \\ 
        \hline $8$ & $1$ & $1$ & $-1$ & $-1$ & 4.25 & $118.2^\circ$ \\ 
        \hline $9$ & $1$ & $1$ & $-1$ & $0$ & 4.25 & $-118.2^\circ$ \\ 
        \hline $10$ & $1$ & $1$ & $0$ & $-1$ & 4.25 & $-118.2^\circ$ \\ 
        \hline $11$ & $1$ & $1$ & $0$ & $1$ & 4.25 & $118.2^\circ$ \\ 
        \hline $12$ & $1$ & $1$ & $1$ & $0$ & 4.25 & $118.2^\circ$ \\ 
        \hline $13$ & $1$ & $1$ & $1$ & $1$ & 4.25 & $-118.2^\circ$ \\ 
        \hline $14$ & $1$ & $2$ & $-1$ & $0$ & 3.51 & $-48.6^\circ$ \\ 
        \hline $15$ & $2$ & $1$ & $0$ & $-1$ & 3.51 & $48.6^\circ$ \\ 
        \hline $16$ & $2$ & $1$ & $0$ & $0$ & 3.51 & $48.6^\circ$ \\ 
        \hline $17$ & $1$ & $2$ & $0$ & $0$ & 3.51 & $-48.6^\circ$ \\ 
        \hline $18$ & $1$ & $2$ & $0$ & $1$ & 3.51 & $-48.6^\circ$ \\ 
        \hline $19$ & $2$ & $1$ & $1$ & $0$ & 3.51 & $48.6^\circ$ \\ 
        \hline $20$ & $2$ & $1$ & $-1$ & $-1$ & 1.76 & $-131.4^\circ$ \\ 
        \hline $21$ & $1$ & $2$ & $-1$ & $-1$ & 1.76 & $131.4^\circ$ \\ 
        \hline $22$ & $1$ & $2$ & $-1$ & $1$ & 1.76 & $131.4^\circ$ \\ 
        \hline $23$ & $2$ & $1$ & $1$ & $-1$ & 1.76 & $-131.4^\circ$ \\ 
        \hline $24$ & $2$ & $1$ & $1$ & $1$ & 1.76 & $-131.4^\circ$ \\ 
        \hline $25$ & $1$ & $2$ & $1$ & $1$ & 1.76 & $131.4^\circ$ \\ 
        \hline $26$ & $1$ & $1$ & $0$ & $0$ & 0.93 & $180.0^\circ$ \\ 
        \hline
    \end{tabular}
    \end{center}
    \caption{
        Parameters of the single valley tight-binding model with $\theta = 3.65^\circ$ and $\varepsilon_D = 15\,\rm meV$. The moir\'e potential strength in this calculation is chosen to be $\tilde{v} = 13\rm\,meV$ and $w = 15\,\rm meV$.
    }
    \label{tab:parameters2}
\end{table}

\section{Projected Coulomb interactions}\label{app:interaction-strength}
Except for the tight-binding parameters shown in the previous section, another output of \textsc{Wannier90} software is the unitary gauge transformation $U_{\alpha, n}(\vk)$, which rotates the single valley eigenstates with the form of Eq.~(\ref{eqn:bloch-eigenstate}), into the Bloch states that correspond to the two local orbitals:
\begin{align}
    \psi_{\alpha\vk}(\mathbf{r}, \ell) &= \frac{1}{\sqrt{\Omega_{\rm tot}}}\sum_{n={1,2}}U_{\alpha, n}(\vk)\sum_{\mathbf{Q} \in \mathcal{Q}_\ell} u_{\mathbf{Q}, \alpha}(\vk)e^{i(\vk - \mathbf{Q})\cdot \mathbf{r}} = \frac{1}{\sqrt{\Omega_{\rm tot}}}\sum_{\mathbf{Q} \in \mathcal{Q}_\ell} \tilde{u}_{\mathbf{Q}, \alpha}(\vk)e^{i(\vk - \mathbf{Q})\cdot \mathbf{r}}\,,\\
    \tilde{u}_{\mathbf{Q},\alpha}(\vk) &= \sum_{n=1,2}U_{\alpha, n}(\vk) u_{\mathbf{Q},n}(\vk)\,.\label{eqn:bloch-orb}
\end{align}
In general, the interacting Hamiltonian can be written as the following form:
\begin{align}
    H_I &= \frac{1}{2\Omega_{\rm tot}}\sum_{\vk\vk'\vq}\sum_{\alpha\beta\alpha'\beta'\sigma\sigma'}U_{\alpha\beta\alpha'\beta'}^{(\sigma\sigma')}(\vq;\vk,\vk') c^\dagger_{\vk+\vq \alpha \sigma} c_{\vk \beta \sigma} c^\dagger_{\vk' - \vq \alpha' \sigma'} c_{\vk' \beta' \sigma'}\,,\\
    U^{(\sigma \sigma')}_{\alpha\beta \alpha' \beta'}(\vq;\vk,\vk') &= \sum_{\mathbf{G} \in \mathcal{Q}_0}V(\vq + \mathbf{G}) \sum_{\mathbf{Q}\in\mathcal{Q}_\pm}\tilde{u}^{(\sigma)*}_{\mathbf{Q}, \alpha}(\vk + \vq + \mathbf{G}) \tilde{u}^{(\sigma)}_{\mathbf{Q}, \beta}(\vk)\sum_{\mathbf{Q} \in \mathcal{Q}_\pm} \tilde{u}^{(\sigma')*}_{\mathbf{Q}', \alpha'}(\vk' - \vq - \mathbf{G}) \tilde{u}^{(\sigma')}_{\mathbf{Q}',\beta'}(\vk')\,.
\end{align}
Here $\tilde{u}^{(\uparrow)}_{\mathbf{Q},\alpha}(\vk)$ are the Bloch state orbital wave functions shown in Eq.~(\ref{eqn:bloch-orb}), and $\tilde{u}^{(\downarrow)}_{\mathbf{Q},\alpha}(\vk)$ are generate from $\tilde{u}^{(\uparrow)}_{\mathbf{Q},\alpha}(\vk)$ through time-reversal transformation.
The Fourier transformation of the screened Coulomb interaction is given by the form of $V(\vq) = (\xi e^2/4\varepsilon_0\varepsilon) \tanh(\xi q /2)/(\xi q/2)$, where $\xi$ is the gate distance, and $\varepsilon$ is the relative dielectric constant.
This interaction vertex $U^{(\sigma\sigma')}_{\alpha\beta\alpha'\beta'}(\vq;\vk,\vk')$ can be Fourier transformed into the real space orbitals as:
\begin{align}
    H_I =& \frac12 \sum_{\mathbf{R}\mathbf{d}\mathbf{d}'\mathbf{R}_0}\sum_{\alpha\beta\alpha'\beta'\sigma\sigma'}\tilde{U}^{(\sigma\sigma')}_{\alpha\beta\alpha'\beta'} (\mathbf{R};\mathbf{d},\mathbf{d}')c^\dagger_{\mathbf{R}+\mathbf{d}+\mathbf{R}_0\alpha\sigma}c_{\mathbf{d} + \mathbf{R}_0\beta\sigma} c^\dagger_{\mathbf{d}' + \mathbf{R}_0\alpha'\sigma'} c_{\mathbf{R}_0\beta'\sigma'}\,,\\
    \tilde{U}^{(\sigma\sigma')}_{\alpha\beta\alpha'\beta'}(\mathbf{R;\mathbf{d},\mathbf{d}'}) =& \frac{1}{N^2\Omega_{\rm tot}}\sum_{\vk\vk'\vq}U_{\alpha\beta\alpha'\beta'}^{(\sigma\sigma')}(\vq;\vk,\vk') e^{i\vq \cdot (\mathbf{R} + \mathbf{d} - \mathbf{d}')}e^{i\vk\cdot \mathbf{d}}e^{i\vk'\cdot \mathbf{d}'}\,.
\end{align}

We computed the value of the projected screened Coulomb interaction with gate distance $\xi = 3\,\rm nm$ at $\mathbf{R} = \mathbf{d} = \mathbf{d}' = \mathbf{0}$ (on-site interaction) for both orbitals with different model parameters.
Using the parameters $(\tilde{v}, w, \varepsilon_D, \theta) = (9{\rm \,meV}, 18{\rm \,meV}, 3{\rm\,meV}, 1.43^\circ)$, we obtain the interaction strengths as
$\varepsilon U_1 = 182.25\,\rm meV$ and $\varepsilon U_2 = 154.35\rm\,meV$.
Using the parameters $(\tilde{v}, w, \varepsilon_D, \theta) = (13{\rm \,meV}, 15{\rm \,meV}, 15{\rm\,meV}, 3.65^\circ)$, we obtain the interaction strength $\varepsilon U_1 = 520.5\rm\,meV$ and $\varepsilon U_2 = 468.15\rm\,meV$.

Since the tWSe$_2$ sample is sandwiched between h-boron-nitride substrates, the reasonable estimation for the effective dielectric function should be in between $\varepsilon_{\rm hBN} \approx 6$ and $\varepsilon_{\rm WSe_2} \approx 16$ \cite{Laturia2018}, which will provide the estimation values for on-site interactions upper and lower limits, respectively.
In the case with twisting angle $\theta = 1.43^\circ$, the interaction strength will be $10 \sim 30\,\rm meV$, while in the case with twisting angle $\theta = 3.65^\circ$, the interaction strength is around $30\sim 85\,\rm meV$.

\section{Extended Dynamical mean-field theory}

In this section, we describe the numerical method used to solve the many-body Hamiltonian. As discussed in the main text, when a finite potential difference is applied, the filling factors of the two orbitals in the effective model differ significantly. 
One orbital is nearly half-filled, while the other is not. 
Consequently, despite similar on-site interaction strengths, the two orbitals exhibit distinct localization tendencies. 
This difference creates a parameter regime analogous to that of a heavy-fermion model.

To simplify the calculation, we explicitly
 include the emergent RKKY exchange interactions between the local moments. 
By tuning the strength of the RKKY interaction, a Kondo destruction quantum critical point can be realized.
This QCP arisesin the framework of EDMFT, in which the single-particle Green's function, the spin susceptibility and the local spin moment expectation values are solved in a self-consistent manner \cite{Hu-Natphys2024,Hu2022extended}.
More precisely, we embed the $f$ orbital into a fermion bath $G^{-1}_0(i\omega)$ and a bosonic bath $\chi^{-1}_0(i\omega)$ to mimic the effect of the hybridization with conduction electrons and the RKKY exchange.
The $f$ orbital Green's function, self energy $\Sigma(i\omega)$ and the fermionic bath $G^{-1}_0(i\omega)$ are determined via the following self-consistent condition:
\begin{align}
     G_{\rm loc}(i\omega) &=\frac{1}{N}\sum_{\vp} \frac{1}{i\omega - \epsilon_f - \Sigma(i\omega) - \frac{V_{\rm hyb}^2}{i\omega - \epsilon_\vp}} \,, \\
     \Sigma(i\omega) &= G^{-1}_0(i\omega) - G_{\rm cl}^{-1}(i\omega) \,,\\
     G_{\rm cl}(i\omega) & = G_{\rm loc}(i\omega)\,.
\end{align}
Similarly, the spin susceptibility and the bosonic bath $\chi_0^{-1}(i\omega)$ are also determined by another set of self-consistent conditions:
\begin{align}
    \chi_{\rm cl}^{-1}(i\omega) &= \chi_0^{-1}(i\omega) + M(i\omega) \,,\\
    \chi_{\rm loc}(i\omega) &= \frac{1}{N}\sum_{\vq} \frac{1}{I_{\vq} + M(i\omega)}\,,\\
    \chi_{\rm loc}(i\omega) &= \chi_{\rm cl}(i\omega)\,,
\end{align}
in which the cluster Green's function $G_{\rm cl}(i\omega)$ and the susceptibility $\chi_{\rm cl}(i\omega)$ are solved exactly accounting for the effects of the fermionic and bosonic baths.
$I_\vq$ is the Fourier transformation of the RKKY interaction, which controls the controls the competition between the Kondo screening and the tendency towards ordering states of the local moments.
In addition to the cluster Green's function and susceptibility, the local spin moment expectation value can also be evaluated from the quantum Monte Carlo simulation.
Therefore, by solving the self-consistent EDMFT equations with different RKKY interaction strengths and at different temperatures, one can determine the critical value of the exchange strength $I_c$, and ordering transition temperature $T_{\rm order}$.
We also note that the quantum Monte Carlo simulation is conducted using imaginary frequencies, requiring an additional analytical continuation step to obtain the spectral function, which is in general a difficult procedure.
Methods that work directly in the real frequency are necessary to properly address this issue, and we leave it for future study.

To allow for the development of unconventional pairing, we adopt a cluster version of EDMFT (C-EDMFT), 
with the cluster containing two sites of $f$-orbitals that are coupled to each other by the antiferromagnetic RKKY interaction. The details of the approach are given in the literature~\cite{Hu2021-sc,Pixley-c-edmft2015}.
In the following, we focus on the spin-singlet pairing that is driven by the antiferromagnetic RKKY interaction. 
The static pairing susceptibility (at zero wavevector), $\chi_{SC}(T)$, involves a form factor, which,
for our case of a triangular lattice, is given by 
$f(\vk) = \cos k_1 + \cos k_2 + \cos k_3$, where $k_i = \vk_i \cdot \mathbf{a}_i$, with $\mathbf{a}_3 = -\mathbf{a}_1-\mathbf{a}_2$. 
The pairing susceptibility is determined by a Bethe Salpeter equation, in terms of 
a bare particle-particle bubble and a irreducible paring vertex function~\cite{Hu2021-sc}. The Brillouin zone is effectively divided into two distinct regions. For a triangular lattice, we select the cluster momenta as $\vK_+ =(0, 0)$ and $\vK_- = (\pm) \frac{1}{3}(\mathbf{b}_1 + \mathbf{b}_2)$, and $\vK$($\vK'$) represents either $\vK_{\pm}$. 
The superconducting phase is identified by the divergence of 
$\chi_{SC}(T)$, which specifies the superconducting transition temperature $T_c$.

\section{Details of the Phase diagram}

In this section, we expand further on the phase diagrams [Figs.~4(a) and 4(b)]  presented in the main text and discuss its context within the broader window of model parameters.

\subsection{Different regimes of interaction strength}

We start our discussion from a two-orbital Hubbard model. 
The interaction effects in these two orbitals can be very different.
The low-energy physics in correlated systems depends on the relative strength of the interaction when compared to various energy scales. 
Briefly speaking, there are four possible cases.
(i) When the interaction strength $U$ is infinitesimally small compared to the width of the flat band crossing the Fermi level (the active band), the low-energy physics is mostly described by a weakly interacting Fermi liquid theory.
(ii) When $U$ is comparable to the active band width but still rather small compare to the band gap, the interaction effect will dominate, although the inter-band effect is still considered as irrelevant.
For example, fractional Chern insulator states are usually considered in this regime.
(iii) When the interaction is further increased to be comparable with the overall band widths of both the two top bands, a multi-orbital description becomes necessary. 
(iv) When the interaction is much stronger than the kinetic energy of the two top-most bands, the two-orbital description itself may fail.
This typically requires the value of $U$ to be more than $\sim 2$ times larger than the band widths of these bands considered together. 
Different interaction strength regimes are summarized in Fig.~\ref{fig:energy-scales}.

\begin{figure}
    \centering
    \includegraphics[width=0.5\linewidth]{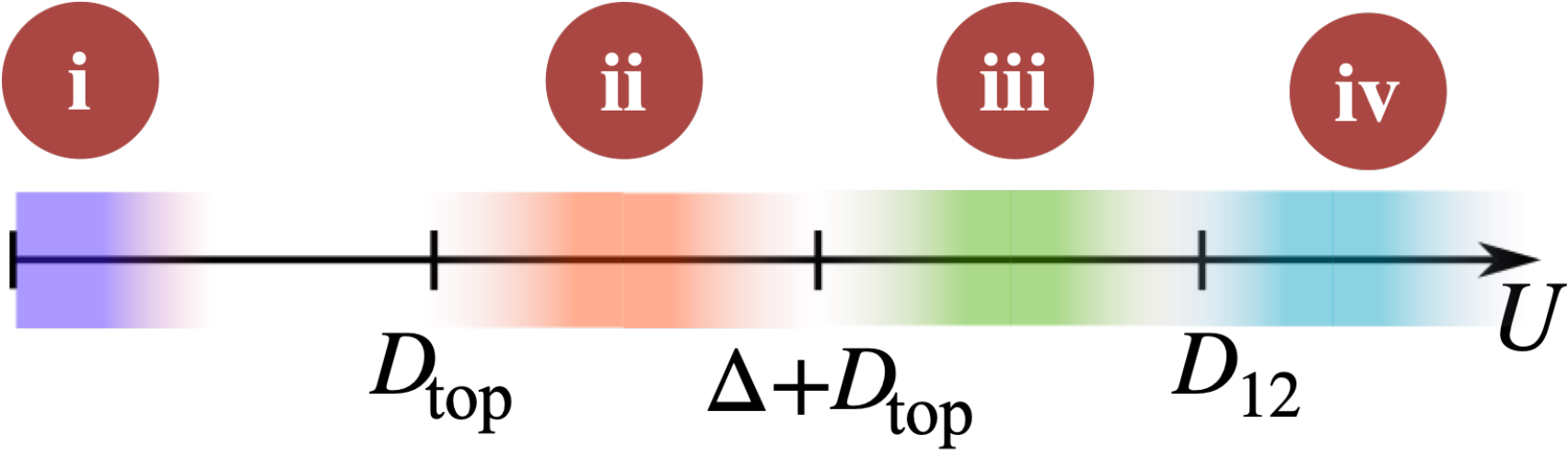}
    \caption{Schematic portrayal of the four interaction regimes. Here $D_{12} \sim D_{\rm top} + D_{\rm bottom} + \Delta$, in which $D_{\rm top}~(D_{\rm bottom})$ stands for the band width of the top (bottom) band, and $\Delta$ is the band gap.}
    \label{fig:energy-scales}
\end{figure}

\subsection{Filling factor and correlation effects}

In multi-orbital interacting systems, the orbital with the smaller band width is more likely to be affected by the on-site interaction.
Additionally, the interaction effect will also be affected by the filling factor of each orbital \cite{Kotliar1986, Hassan2010slave}.
In Fig.~\ref{fig:z-vs-x}, we provide the quasiparticle weight $Z$ as a function of filling factor $\nu = 1+x$ in a single-orbital Hubbard model by $U(1)$ slave spin approach \cite{Yu2012U1slave}.
It can be shown that the orbital with filling factor closer to half-filling will experience a stronger correlation effect.
In this situation, the effective description of the system is the Anderson lattice model.
The orbitals with stronger correlation effects could be treated as the $f$ orbital, and the other can be treated as a conduction band.

\begin{figure}
    \centering
    \includegraphics[width=0.5\linewidth]{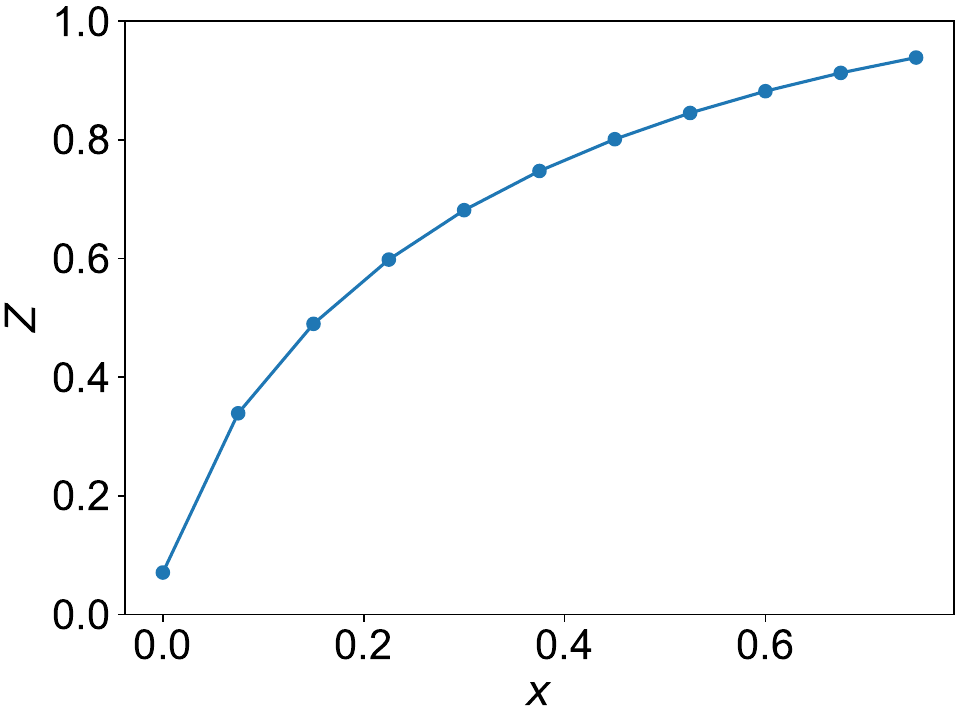}
    \caption{Doping level $x = \nu - 1$ versus quasiparticle weight $Z$ for one-band Hubbard model solved via $U(1)$ slave spin method. Here we use a 2D square lattice, with interaction strength $U/t = 12.5$.}
    \label{fig:z-vs-x}
\end{figure}

\subsection{Role of hybridization}

The hybridization between these two orbitals $V_{\rm hyb}$ plays an important role in the quantum phases in this Hamiltonian.
The Kondo coupling between the conduction electrons and the $f$ orbital spin moment is induced by the hybridization through a second order perturbation, which has the form of $J_K \sim V_{\rm hyb}^2\left(\frac{1}{\epsilon_f} + \frac{1}{-\epsilon_f + U}\right)$.
The energy scale of the Kondo screening strength is depicted by the Kondo temperature $T_K \sim \rho_0^{-1}\exp(-1/J_K\rho_0)$.
In the mean time, the conduction electrons can also mediate the RKKY exchange interaction between the $f$-orbital spin moments in different unit cells, whose strength is controlled by $I \sim \rho_0 J_K^2$, in which $\rho_0$ is the density of states of the conduction electrons at Fermi level.

Since the Kondo temperature $T_K$ and the RKKY exchange strength are controlled by the $V_{\rm hyb}$ in a different manner, changing the strength of $V_{\rm hyb}$ will also change the relative strength between these two quantities, and therefore control the competition between them.
As depicted in Fig.~4(a), with a weaker hybridization, the RKKY interaction dominates, and the $f$-orbital spin moments tend to form magnetic ordered state. With a stronger hybridization, the effect of Kondo screening also becomes stronger, and could drive the system across a quantum critical point into a Fermi liquid phase.
The spin moment fluctuation near the quantum critical point can lead to superconducting instability at low temperatures.

In the C-EDMFT treatment of the Anderson lattice model, the RKKY interaction is considered as an independent parameter to the hybridization strength, as shown in Eq.~(2) in the main text.
The quantum phase transition can also be achieved by tuning the value of $I$ with fixed $V_{\rm hyb}$, as demonstrated in Fig.~4(b) in the main text.
This is more convenient in the framework of C-EDMFT technologically.
The phase with stronger $V_{\rm hyb}$ in Fig.~4(a), which corresponds to the Kondo screened phase, is the same phase with weak exchange interaction $I$ in Fig.~4(b).

We also comment on the effect of on-site interaction strength $U$ in the phase diagram.
As mentioned in the previous paragraphs, one of the most important energy scales, the Kondo coupling $J_K$, can be influenced by the interaction strength.
Since superconductivity arises from spin fluctuations, the transition temperature $T_c$ may also be affected. 
However, in all our calculations, we find $T_c/T_K$ to be on the order of a few percent across different interaction strengths. 
In our case, the value of $\epsilon_f$ (measured relative to the chemical potential) is relatively small compared to $U$, the effect of $U$ on $J_K$ is not too strong, suggesting that $T_c$ might not be highly sensitive to $U$.

\section{Robustness of the superconductivity based on additional numerical results}\label{sec:app-numerical}

In this section, we present our additional numerical results.
Here, we use the tight-binding model with parameters $(\tilde{v}, w, \varepsilon_D, \theta) = (13{\rm \,meV}, 15{\rm \,meV}, 15{\rm\,meV}, 3.65^\circ)$.
As we have already explained before, the top two moir\'e bands can be Wannierized with these parameters choice.
The band structure and the corresponding orbitals are summarized in Fig.~\ref{fig:wannier-365}.

In addition, using the tight-binding parameters listed in Table \ref{tab:parameters2}, we estimate the average hybridization between the two orbitals to be around $V_{\rm hyb} = 7\rm\,meV$.
We solved this model using C-EDMFT with different interaction strengths $U = 40\,\rm meV$ and $U = 80\,\rm meV$.
Since the interaction strength $U$ is noticeably larger than the value of $\epsilon_f - \mu$, we expect that the Kondo coupling is not very sensitive to the value of $U$.
The numerical simulation shows that the bare Kondo temperature scales in these two cases are given by $k_BT^0_K = 18.43\rm\,meV$ and $k_B T^0_K = 18.36\rm \, meV$, respectively.
The superconducting transition temperature, and the electronic order transition temperature are also solved, which are shown in Fig.~\ref{fig:edmft-large-angle}. 
Similar to the results shown in the main text, $T_c$ is able to approach a few percent of the Kondo temperature, which indeed aligns with the experimental observation.
In addition, since the effective Kondo coupling scale is not very sensitive to the $U$ once it is already in the suitable regime, the transition temperatures are also close to each other, although the interaction strengths are different.
This result also demonstrates the robustness of the mechanism we have proposed for 
superconductivity against interaction parameter tuning.

\begin{figure}
    \centering
    \includegraphics[width=0.7\linewidth]{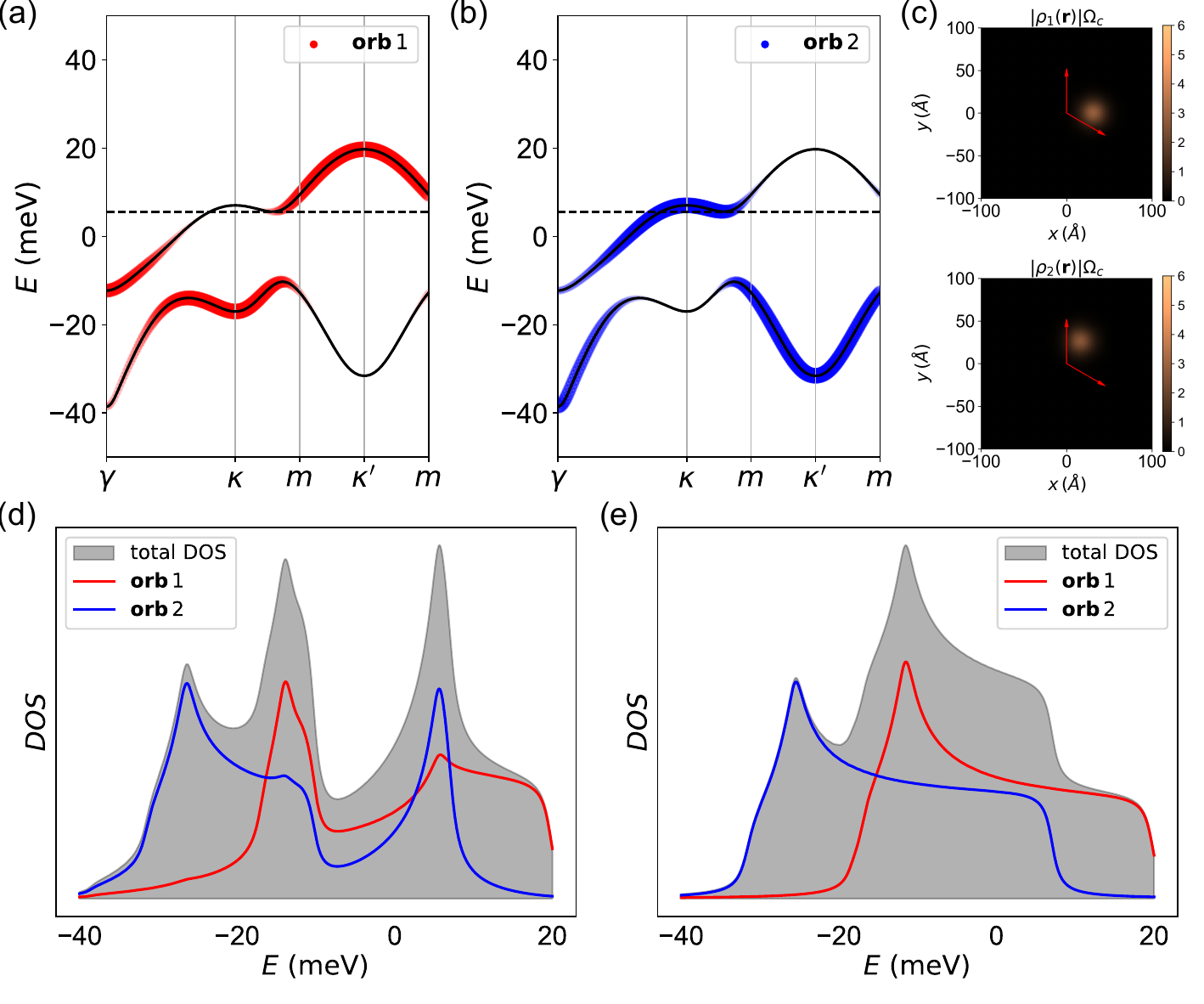}
    \caption{(a-b) The fat band plots of the two Wannier orbitals.
    (c) The real-space density distributions of these two orbitals.
    (d) The density of states or the two-orbital tight-binding model.
    (e) The density of states without the hybridization between the two orbitals.
    The band structure parameters are chosen as $(\tilde{v}, w, \varepsilon_D, \theta) = (13{\rm \,meV}, 15{\rm \,meV}, 15{\rm\,meV}, 3.65^\circ)$.
    }
    \label{fig:wannier-365}
\end{figure}

\begin{figure}
    \centering
    \includegraphics[width=0.7\linewidth]{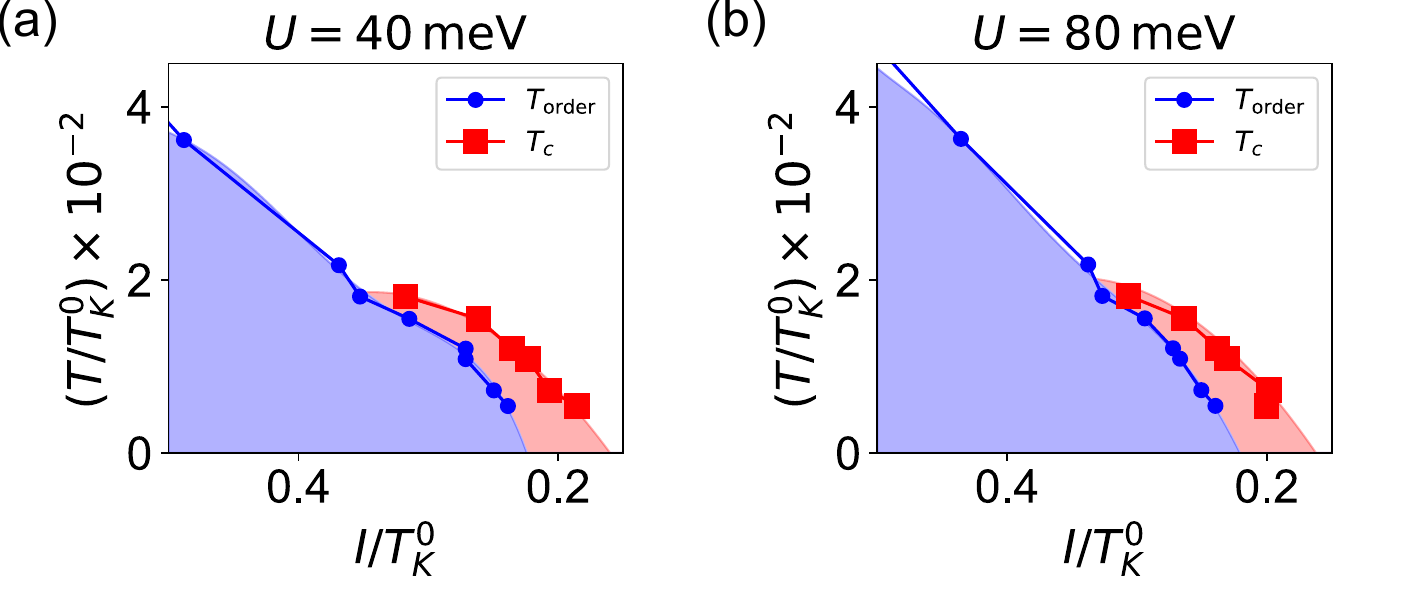}
    \caption{The superconducting transition temperature $T_c$, and the magnetic ordering temperature $T_{\rm order}$ scales as functions of the exchange coupling $I$ with interaction strength $U = 40\,\rm meV$ and $U = 80\,\rm meV$.}
    \label{fig:edmft-large-angle}
\end{figure}

\section{Connection with experiments}

The superconductivity phase in twisted bilayer $\rm WSe_2$ has already been observed in two independent experiments \cite{Xia2024Unconventional, Guo2024Superconductivity}.
The filling factor can be controlled by the gate voltages, and the superconducting state is discovered when the sample is doped by one hole per moir\'e unit cell. 
A correlated phase is also observed in the vicinity of this superconducting phase. 

Experiments have provided evidence for the strongly correlated nature of the normal state.
The resistivity of Landau Fermi liquid behaves like $\rho(T) \approx \rho_0 + AT^2$ at low temperatures, in which the coefficient $A$ is  proportional to the square of effective mass.
In tWSe$_2$, when the filling factor approaches the superconducting regime in the phase diagram, 
the value of $A$ is strongly increased, as shown in Fig.~\ref{fig:experiment-quasiparticle-weight}:
$\sqrt{A}$ increases by more than a factor of $10$. This is highly reminiscent to what is observed 
in quantum critical heavy fermion metals. By comparison with the latter, it
suggests that the normal state of the superconducting 
tWSe$_2$ is in proximity to a quantum critical point,
as we have proposed theoretically.

\begin{figure}
    \centering
    \includegraphics[width=0.5\linewidth]{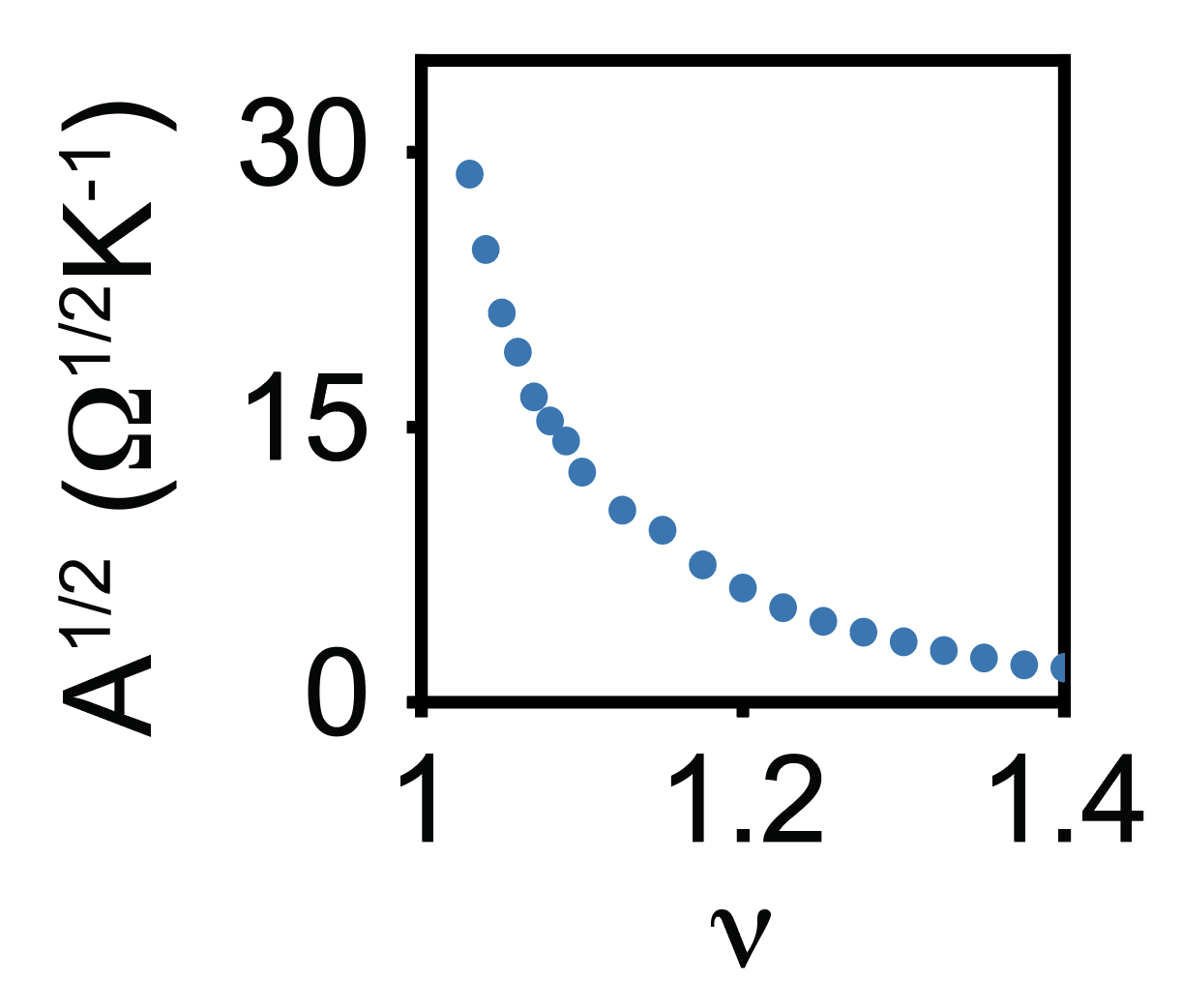}
    \caption{The coefficient $A$ as a function of filling factor in the experimental observation. The quantity $A$ is defined as the coefficient of $T^2$-term in the temperature dependent resistivity in Fermi liquid $\rho(T) \approx \rho_0 + AT^2$. 
    This figure is adapted from Ref.~\cite{Xia2024Unconventional}.
    }
    \label{fig:experiment-quasiparticle-weight}
\end{figure}

\end{document}